\documentclass{emulateapj} 

\begin{document}

\newcommand{\cl}{{RX~J0152-13}}
\newcommand{\ms}{{MS~1054-03}}
\newcommand{\ctw}{{CL~J1226+33}}
\newcommand{\clt}{{CL~1358+62}}
\newcommand{\od}{{CL~0054-27}}
\newcommand{\mst}{{MS~2053-04}}
\newcommand{\zwcl}{{CL~0024+16}}
\newcommand{\koo}{{CL~0016+16}}
\newcommand{\hoga}{{CL~1604+4321}}
\newcommand{\hogb}{{CL~1604+4304}}
\newcommand{\lynx}{{RX~J0849+4452}}
\newcommand{\mei}{{RX~J0910+5422}}
\newcommand{\jpb}{{RX~J1252-2927}}
\newcommand{\ecla}{{ECL~J1040-11}}
\newcommand{\eclb}{{ECL~J1054-11}}
\newcommand{\eclc}{{ECL~J1054-12}}
\newcommand{\ecld}{{ECL~J1216-12}}
\newcommand{\eclm}{{ECL~J1232-12}}

\newcommand{\g}{$g_{475}$}
\newcommand{\B}{$B_{435}$}
\newcommand{\V}{$V_{606}$}
\newcommand{\ra}{$r_{625}$}
\newcommand{\ia}{$i_{775}$}
\newcommand{\I}{$I_{814}$}
\newcommand{\z}{$z_{850}$}

\newcommand{\ntot}{675}
\newcommand{\nlum}{453}
\newcommand{\nmass}{441}
\newcommand{\nmag}{487}
\newcommand{\lowzlim}{210}
\newcommand{\lowzlime}{47}
\newcommand{\lowzlims}{83}

\newcommand{\etal}{{\em et~al.\,}}

\title{The Ellipticities of Cluster Early-type Galaxies from
  $z\sim$1 to $z\sim 0$: No Evolution in the Overall
  Distribution of Bulge-to-Disk Ratios
  \altaffilmark{1}\altaffilmark{2}\altaffilmark{3}}

\altaffiltext{1}{Based on observations with the NASA/ESA Hubble Space
  Telescope, obtained at the Space Telescope Science Institute, which
  is operated by the Association of Universities for Research in
  Astronomy, Inc. under NASA contract No. NAS5-26555.} 
\altaffiltext{2}{Some of the data presented herein were obtained at
  the W.M. Keck Observatory, which is operated as a scientific
  partnership among the California Institute of Technology, the
  University of California and the National Aeronautics and Space
  Administration. The Observatory was made possible by the generous
  financial support of the W.M. Keck Foundation.}
\altaffiltext{3}{This paper includes data gathered with the 6.5 meter
  Magellan Telescopes located at Las Campanas Observatory, Chile.}

\author{B. P. Holden\altaffilmark{4}}

\author{M. Franx\altaffilmark{5}}

\author{G. D. Illingworth\altaffilmark{4}}

\author{M. Postman\altaffilmark{6}}

\author{A. van der Wel\altaffilmark{7}}

\author{D.~D. Kelson\altaffilmark{8}}

\author{J.~P. Blakeslee\altaffilmark{9}}

\author{H. Ford\altaffilmark{7}}

\author{R. Demarco\altaffilmark{7}}

\author{S. Mei\altaffilmark{7,10,11}}

\altaffiltext{4}{UCO/Lick Observatories, University of California,
  Santa Cruz, 95065; holden@ucolick.org; gdi@ucolick.org}

\altaffiltext{5}{Leiden Observatory, University of Leiden, P.O.Box
  9513, 2300 RA, Leiden, The Netherlands; franx@strw.leidenuniv.nl}

\altaffiltext{6}{Space Telescope Science Institute, Baltimore, MD 21218;
  postman@stsci.edu}

\altaffiltext{7}{Department of Physics \& Astronomy, Johns Hopkins University,
  Baltimore, MD 21218; wel@pha.jhu.edu; 
  ford@pha.jhu.edu; demarco@pha.jhu.edu}

\altaffiltext{8}{Observatories of the Carnegie Institution of
  Washington, Pasadena, CA, 91101; kelson@ociw.edu}

\altaffiltext{9}{Herzberg Institute of Astrophysics, National Research
  Council of Canada, Victoria, BC V9E2E7, Canada; john.blakeslee@nrc.ca} 

\altaffiltext{10}{University of Paris Denis Diderot, 75205 Paris Cedex
  13, France} 

\altaffiltext{11}{GEPI, Observatoire de Paris, Section de Meudon, 5
  Place J. Janssen, 92195 Meudon Cedex, France; simona.mei@obspm.fr}

\shorttitle{No Ellipticity Evolution for Cluster Early-Type Galaxies}

\begin{abstract} 

  We have compiled a sample of early-type cluster galaxies from $0 < z
  < 1.3$ and measured the evolution of their ellipticity
  distributions. Our sample contains \nmag\ galaxies in 17 $z>0.3$
  clusters with high-quality space-based imaging and a comparable
  sample of \lowzlim\ galaxies in 10 clusters at $z<0.05$. We select
  early-type galaxies (elliptical and S0 galaxies) that fall within
  the cluster $R_{200}$, and which lie on the red-sequence in the
  magnitude range $-19.3 > M_B > -21$, after correcting for luminosity
  evolution as measured by the fundamental plane. Our ellipticity
  measurements are made in a consistent manner over our whole
  sample. We perform extensive simulations to quantify the systematic
  and statistical errors, and find that it is crucial to use
  point-spread function (PSF)-corrected model fits; determinations of
  the ellipticity from {\em Hubble Space Telescope} (HST) image data
  that do not account for the PSF ``blurring'' are systematically and
  significantly biased to rounder ellipticities at redshifts $z>0.3$.
  We find that neither the median ellipticity, nor the shape of the
  ellipticity distribution of cluster early-type galaxies evolves with
  redshift from $z\sim 0$ to $z > 1$ (i.e., over the last
  $\sim8$Gyr). The median ellipticity at $z>0.3$ is statistically
  identical with that at $z<0.05$, being higher by only 0.01 $\pm$
  0.02 or 3 $\pm$ 6\%, while the distribution of ellipticities at
  $z>0.3$ agrees with the shape of the $z<0.05$ distribution at the
  1-2\% level (i.e., the probability that they are drawn from the same
  distribution is 98-99\%). These results are strongly suggestive of
  an unchanging overall bulge-to-disk ratio distribution for cluster
  early-type galaxies over the last $\sim8$Gyr from $z\sim1$ to
  $z\sim0$.  This result contrasts with that from visual
  classifications which show that the fraction of
  morphologically-selected disk-dominated early-type galaxies, or S0s,
  is significantly lower at $z>0.4$ than at $z\sim0$.  We find that
  the median disk-dominated early-type, or S0, galaxy has a somewhat
  higher ellipticity at $z>0.3$, suggesting that rounder S0s are being
  assigned as ellipticals. Taking the ellipticity measurements and
  assuming, as in all previous studies, that the intrinsic ellipticity
  distribution of both elliptical and S0 galaxies remains constant,
  then we conclude from the lack of evolution in the observed
  early-type ellipticity distribution that the relative fractions of
  ellipticals and S0s do not evolve from $z\sim1$ to $z=0$ for a
  red-sequence selected samples of galaxies in the cores of clusters
  of galaxies.
 
\end{abstract}

\keywords{galaxies: clusters: general --- galaxies: elliptical and
lenticular, cD, --- galaxies: evolution --- galaxies: fundamental
parameters --- galaxies: photometry --- clusters: individual: CL~J1226.9+3332 }

\section{Introduction}

S0 galaxies are enigmatic objects whose formation and evolution is
still not well understood.  Originally, these galaxies were postulated
to exist as a transition class between the elliptical and spiral
sequence \citep{hubble1936}.  Both elliptical and S0 galaxies lack
spiral arms or major dust features.  However, early work on the
ellipticity distributions of galaxies showed that S0 galaxies were
disk-dominated systems with ellipticity distributions that differed from
ellipticals, being more similar to spiral galaxies in their intrinsic
shapes \citep[{\it e.g.},][]{rood67,sandage1970}.  The early work on the
morphology-density relation by \citet{dressler1980b} emphasized that
S0s and ellipticals both occur with higher frequency in higher
density environments, while the pioneering studies of \citet{bo1984}
and \citet{dg92} began to provide hints about how the early-type
population of elliptical and S0 galaxies might evolve out to redshifts
around 0.5 and earlier.

However, it was not until {\em HST} allowed comprehensive
high-resolution imaging of distant clusters of galaxies that
\citet{dressler97}, for example, and others began to show directly
that the S0 fraction was changing at high redshift.  This early work,
along with more recent studies \citep[e.g.,][]{postman2005,desai2007}
found smaller S0 galaxy fractions in clusters of galaxies at higher
redshifts, $z>0.3-0.4$.  The implication is that the S0 galaxy
population forms with different time-scales and later than the
elliptical population.  \citet{fasano2000}, \citet{smith2005},
\citet{postman2005}, \citet{poggianti2006}, \citet{desai2007}, and,
most recently, \citet{wilman2008} all find that the majority of
evolution occurs since $z\sim0.4$, i.e., in the past 4 Gyrs, and above
those redshifts there is little or no evolution in the early-type
galaxy fraction in clusters of galaxies out to $z\sim1$
\citep[see][for a different point of view]{smith2005}.  Since S0 and
elliptical galaxies have different ellipticity distributions and
bulge-to-disk distributions, two simple (and related) predictions that
can be drawn from the observed changes in S0 fraction with redshift is
that both the mean bulge-to-disk ratio of early-type cluster galaxies
and the ellipticity distribution should have changed over relatively
recent epochs (since $z\sim 0.4$).

The evidence for evolution in the S0 population of clusters rests
primarily on morphological classifications of galaxies.  The
separation of the early-type galaxy population into S0 and elliptical
galaxies has long been recognized as being a challenging task (see,
for example, \citealt{andreon98} for a discussion of the systematic errors in
classification and how misclassification mimics evolution.) S0
galaxies are defined as multicomponent disk galaxies, while
ellipticals are defined as single component systems.  However, a
number of quantitative studies found that a substantial fraction of
ellipticals have ``disky'' isophotes \citep[see][for a
summary]{kormendy1989} in contrast with this definition.  The analysis
of \citet{rix1990} exemplifies the challenge of establishing the
relative contributions of elliptical and S0 galaxies.  They showed,
for a $z\sim0$ galaxy, that detecting a disk component in a spheroidal
galaxy is increasingly difficult as the disk becomes more face-on in
projection.  \citet{rix1990} found that a disk containing 20\% of the
total light of a galaxy is impossible to detect over half of the range
of $\cos(i)$ where $i$ is the inclination angle.  Specifically, they
note ``since the cos $i$ axis can be interpreted as a probability
axis, this implies that 50\% of all disks with $L_D/L_B < 0.25$ cannot
be detected by photometric means.''  These, however, results for a
single galaxy.  For an ensemble of galaxies, the bulge-to-disk ratio
distribution can be constrained by the intrinsic ellipticity
distribution, with the average observed ellipticity being directly
related to the intrinsic ellipticity \citep{binneym1998}.  Therefore,
examining the distribution of galaxy ellipticities provides a direct
measure of the evolution in the distribution of the bulge-to-disk
ratios of that galaxy population.

\citet{jf94} investigated the ellipticity distribution of elliptical
and S0 galaxies in the Coma cluster, and suggested that elliptical and
S0 galaxies were not distinct classes but were part of a continuum of
objects of varying bulge-to-disk ratios.  This result was given
additional support by a recent study by \citet{krajnovic2008} who
found that 69\% of elliptical galaxies have multiple kinematic
components, generally disk-like components, while 92\% of S0 galaxies
have disk-like components.  These results give emphasis to the view
that elliptical and S0 galaxies form a continuous distribution of disk
fraction as opposed to two distinct classes. In particular,
\citet{jf94} constructed a model of the ellipticity distribution of
elliptical and S0 galaxies using a continuum of bulge and disk
components viewed from a variety of angles.  \citet{jf94} found a
deficit of round S0 galaxies, in contrast with what was expected from
their model, suggesting that some face-on S0 galaxies had been
classified as ellipticals.  \citet{blakeslee2005} and \citet{mei2005a}
also found a lack of round S0 galaxies in three $z\sim 1$ clusters
of galaxies when compared to what was expected for a disk population
viewed at a variety of angles.  These three studies illustrate a
potential bias in the visual classification of galaxies, namely that
nearly face-on S0 galaxies galaxies may be incorrectly classified as
ellipticals. A number of studies suggest that quantitative
measurements do not suffer the same orientation bias as visual
classifications \citep{blakeslee2005,vanderwel2008b}, but robust,
bias-free galaxy classification, either visual or quantitative, still
remains an elusive goal.

The importance of the work showing an apparent evolution in the S0
fraction in clusters and the knowledge of potential classification
biases led \citet{dressler97} and \citet{postman2005} to investigate
the ellipticity distributions of the S0 and elliptical galaxies in
their samples.  In general, the authors found that the ellipticity
distributions of S0 and elliptical galaxies show no evolution over the
broad redshift ranges in their samples \citep[][compared clusters
from $z\sim0.25$ to $z\sim1.3$]{postman2005}.  Also, the
ellipticity distributions of elliptical and S0 galaxies differ from
each other, providing evidence for the existence of two distinct
classes of galaxies. However, these previous studies do not use a
consistent measure of the ellipticity as compared with the $z\sim0$
efforts such as \citet{jf94} or \citet{andreon1996}, making a
comparison between these $z>0.2$ samples observed with {\em HST} and
$z\sim 0$ samples observed from the ground difficult. Traditionally,
ellipticities for galaxies at $z\sim0$ were measured by visual
estimates \citep{dressler1980b} or by fitting models to the elliptical
isophotes \citep{jf94,andreon1996}.  At higher redshifts, $z>0.2$, the
ellipticities in \citet{smail1997} or \citet{postman2005} are
determined by the second-order flux-weighted moments of a detection
isophote.

Both the data and the techniques have matured so that we can now
evaluate the ellipticity distribution, as a function of redshift,
quantitatively, and even more importantly, in a consistent way with
minimal systematic error.  We show that some approaches used
previously that could not correct for the point-spread function (PSF)
are probably subject to significant systematic error.  Essentially,
the ``blurring'' effect of the PSF will lead to galaxies being
measured as rounder than they actually are.  We will use a single
consistent and robust approach for measuring the ellipticities at all
redshifts.  The ellipticities will be measured by fitting models
convolved with the PSF to galaxy surface brightness profiles.  This
will eliminate some of the previously-reported uncertainties found
when comparing ground-based imaging data taken under different seeing
conditions \citep[see][for some discussion]{andreon1996}. In addition,
this approach essentially eliminates the systematic error associated
with the PSF ``blurring''.  Our ellipticity measurements will provide an
assessment of the evolution in the distribution of the overall
bulge-to-disk ratio of early-type cluster galaxies from the
present day to redshifts $z\sim 1$.
 
We will discuss how we compiled our samples of early-type cluster
galaxies with morphological classifications in \S \ref{sample}, and
then how we measured their total magnitudes, colors, and ellipticities
in \S \ref{data}.  One of the advantages of our approach of using an
automated measurement technique is that we can simulate the
measurement process.  We discuss this in \S \ref{ellmeas} and in
Appendix A1.  From our measurements, we find no evolution in the
distribution of ellipticities of cluster early-type galaxies, which we
show in \S \ref{elldist}.  In \S \ref{disc}, we discuss the
implications of this result and contrast with previous measurements of
the evolution in the overall distribution of the bulge-to-disk ratio of
cluster early-type galaxy population.  We follow this with a summary and some
discussion of the broader implications in \S \ref{conclusion}.

Throughout this paper, we assume $\Omega_m = 0.27$, $\Omega_{\Lambda}
= 0.73$ and $H_o = 71\ {\rm km\ s^{-1}\ Mpc^{-1}}$.  All $B$
magnitudes we list use the Vega zeropoint ($B_{Vega} = B_{AB} + 0.11$.)
Other magnitudes where given are AB mags.

\section{Sample Selection}
\label{sample}

To carry out our study of the ellipticities of early-type galaxies
from $z\sim1$ to the current epoch, we have assembled a sample of
morphologically-selected, cluster early-type galaxies ranging in
redshifts from $z \sim 0$ to $1.27$.  The clusters used are
tabulated below, as is the source of the morphologies for the galaxies
samples (\S \ref{morph}). The early-type galaxies are chosen to lie
within the projected $R_{200}$, which we use to define cluster membership.
The measurement for each galaxy of its color, magnitude, and 
ellipticity, is described  in  \S \ref{data}.

\subsection{Cluster Sample Selection}

\begin{deluxetable*}{lrrrr}
\tablecolumns{5}
\tablecaption{$z<0.05$ Cluster Summary}
\tablehead{\colhead{Cluster} & \colhead{z}\tablenotemark{a} & 
\colhead{$\sigma$\tablenotemark{a}} & \colhead{$R_{200}$\tablenotemark{b}} 
 & \colhead{\# \ Early-type\tablenotemark{c}}
  \\
\colhead{} & \colhead{} & \colhead{(km s$^{-1}$)} & \colhead{(Mpc)} & \colhead{}  \\}

\startdata
ACO 119 & 0.0440 & 744 & 1.13 & 41 \\ 
ACO 168 & 0.0452 & 524 & 0.80 & 54 \\ 
ACO 194 & 0.0178 & 435 & 0.67 & 9 \\ 
ACO 957 & 0.0440 & 691 & 1.05 & 32 \\ 
ACO 1139 & 0.0383 & 436\tablenotemark{d} & 0.67 & 23 \\ 
ACO 1142 & 0.0353 & 417 & 0.64 & 17 \\ 
ACO 1656 & 0.023 & 1008 & 1.55 & 203 \\
ACO 1983 & 0.0441  & 433 & 0.66 & 18 \\ 
ACO 2040 & 0.0456  & 602 & 0.92 & 50 \\ 
ACO 2063 & 0.0337  & 521 & 0.80 & 50 \\ 
ACO 2151 & 0.0371  & 786 & 1.20 & 58 \\                                               

\enddata
\tablenotetext{a}{The cluster redshift and $\sigma$ from
  \citet{struble99} unless otherwise noted.  }
\tablenotetext{b}{$R_{200}$ is derived from $\sigma$.}
\tablenotetext{c}{The number of galaxies with E, E/S0, S0/E, S0, or S/S0 classifications.}
\tablenotetext{d}{From \citet{poggianti2006}}

\label{lowz_num}
\end{deluxetable*}

Our $z<0.05$ sample is selected from Abell clusters in the Sloan
Digital Sky Survey Fifth Data Release \citep[SDSS DR5]{dr5}. We
summarize this sample in Table \ref{lowz_num}, where we list the Abell
number \citep{aco89} and the redshift of the cluster from
\citet{struble99}.  The last column lists the number of
redshift-selected members with early-type classification, which lie
within the projected $R_{200}$ and are on the $g-r$ red sequence in
the SDSS DR5 imaging.

To determine $R_{200}$, we will use the formula give in
\citet{carlberg97}, or \[ R_{200} = \frac{\sqrt{3}}{10}
\frac{\sigma_1}{H(z)} \] where $\sigma_1$ is the one-dimensional
velocity dispersion and $H(z)$ is the Hubble constant at the redshift
of observation.  \citet{desai2007} uses the relation from
\citet{finn2005}, which is functionally the same, so effectively our
selection radii are similar to \citet{desai2007}.  We selected $2 R_{200}/\pi
$, instead of $R_{200}$, as a galaxy at $R_{200}$ from the
cluster center will be, on average, projected to appear at the
distance $2 R_{200}/\pi$ \citep[see][for example]{limber1960}.  

For the high redshift clusters, we required {\em Hubble Space
  Telescope} ({\em HST}) imaging with the Advanced Camera for Surveys
(ACS) or the Wide Field Planetary Camera 2 (WFPC2) in multiple
filters.  We also required that the morphological classifications were
done in a way consistent with the original
\citet{dressler1980a} and \citet{dressler1980b} work.  For the three clusters
where $2 R_{200}/\pi $ was larger than the field of view over which we
had imaging data \citep[generally the clusters in the sample of
][]{dressler97}, we simply used all available galaxies. We mark those
clusters in Table \ref{summary} with the superscript b.  Below we will
detail the sources for the morphological classifications and how we
measured the total magnitudes, colors, and ellipticities.

We tabulate the $0.3 < z < 1.3$ clusters for which we have early-type
galaxy samples in Table \ref{summary}. In that table, we list the
dispersions, inferred radii, and the final sample sizes.  For \ctw\ we
have compiled a new catalog of members which we will discuss in
Appendix \ref{ctw}.  For \lynx, we compute a new dispersion.  We use
both redshifts from previous work \citep{stanford2001,mei2005b} and
unpublished ones we have recently collected (which we plan to publish
in a future paper.)  There are a total of 18 galaxies in \lynx\ with
redshifts within $R_{200}$.  The biweight center of the distribution
is $z=1.2600 \pm 0.0017$ and the dispersion is 798 $\pm$ 208 km
s$^{-1}$.  The errors for both the redshift and the dispersion are
estimated by bootstrapping the redshift distribution.

For the clusters \cl, \ms, \mst, and \clt, we used the redshift
catalogs from \citet{holden2007}. The faint magnitude limit we adopted
corresponds to the completeness limits for our high redshift samples
\citep{holden2007}. For the remaining clusters, we included all
galaxies that met our magnitude, color, morphology and radial
requirements, rejecting only those galaxies with redshifts outside of
the clusters.  This will mean that some cluster samples will be
contaminated, i.e., some red-sequence early-type galaxies will not be
members.  The level of contamination is discussed below.  Our $z>0.3$
sample consists of \nmag\ early-type galaxies while our $z<0.05$
sample has \lowzlim\ galaxies.

Many of the cluster red-sequence selections we use are tabulated in
other studies.  For those clusters studied by the ACS Instrument
Definition Team, these red-sequence relations are summarized in
\citet{mei2008}.  For the remaining clusters, we derive the early-type
red-sequence from the data.  Whenever possible, we use existing
redshift catalogs to determine the red-sequence of spectroscopic
determined members.  We then accept all galaxies that lie within the
2$\sigma$ of that sequence, rejecting those galaxies that are known
not to be members.

We know that there will be some contamination by field galaxies in our
catalogs.  To measure the level of contamination, we used our
redshift catalogs for \ms\ and \cl\ along with the catalogs of EDisCS
from \citet{halliday2004}.  \cl\ has a known group in the foreground
with colors very similar to that of the cluster \citep[see][for a
discussion]{holden2005b}.  Nonetheless, when we consider all
early-type galaxies that lie on the red-sequence for \ms\ and \cl,
regardless of redshift, we find a 3 $\pm$ 1\% contamination rate of
nonmembers.  When we examine the less extremely rich sample of the
EDisCS, we find a contamination fraction of 10 $\pm$\ 3\%, which is
likely more representative of the typical clusters in our sample.  The
small size of this contamination means we need not compute a
statistical background correction for the cluster red-sequence.

\begin{deluxetable*}{lrrrrrrr}
\tablecolumns{7}
\tablecaption{Summary of $0.3 < z < 1.3$ Cluster Data}
\tablehead{\colhead{Cluster} & \colhead{z} & \colhead{R.A.} &
  \colhead{Dec.} & \colhead{Obs. Filters} & \colhead{$\sigma$} &
  \colhead{$R_{200}$} & \colhead{\# Early-type \tablenotemark{a}} \\
\colhead{} & \colhead{} & \colhead{} & \colhead{}  & \colhead{} & 
\colhead{(km s$^{-1}$)} & \colhead{(Mpc)} & \colhead{}  \\
}
\startdata
\clt & 0.328\tablenotemark{c}& 13 59 50.6 & +62 59 05 &$V_{606}\ I_{814}$ &
1027$^{+51}_{-45}$\tablenotemark{c} & 1.4 &  67 \\
\zwcl & 0.395 & 00 26 35.7 & +17 09 46 & $r_{625}\ i_{775}\ z_{850}$ & 650$^{+50}_{-50}$ & 0.83\tablenotemark{b} &  26 \\                     
\koo & 0.541 & 00 18 33.5 & +16 26 14 &$i_{775}\ z_{850}$  &
1234$^{+128}_{-128}$ \tablenotemark{d} & 1.46\tablenotemark{b} & 42 \\   
\eclm & 0.541\tablenotemark{e} & 12 32 30.3 & -12 50 36 & $V$ $I$, $I_{814}$ &
$1080^{+119}_{-99}$\tablenotemark{e} & 1.28 & 21 \\
\od & 0.56 & 00 56 56.9 & -27 40 30 &$V_{606}\ I_{814}$   & 1180\tablenotemark{f}  & 1.38\tablenotemark{b} &  19 \\           
\mst & 0.587 & 20 56 21.3 & -04 37 51 &$V_{606}\ I_{814}$  &
865$^{+71}_{-71}$ \tablenotemark{g} & 1.00 &  31 \\    

\eclb & 0.697\tablenotemark{e} & 10 54 24.4 & -11 46 19 & $V$ $I$, $I_{814}$ &
$589^{+78}_{-70}$ \tablenotemark{e} & 0.64 & 5 \\
\ecla & 0.704\tablenotemark{e} & 10 40 40.3 & -11 56 04 & $V$ $I$, $I_{814}$ &
$418^{+55}_{-45}$ \tablenotemark{e} & 0.45 & 2 \\
\eclc & 0.750\tablenotemark{e} & 10 54 43.5 & -12 45 52 & $V$ $I$, $I_{814}$ &
$504^{+113}_{-65}$ \tablenotemark{e} & 0.53 & 13 \\
\ecld & 0.794\tablenotemark{e} & 12 16 45.3 & -12 01 18 & $V$ $I$, $I_{814}$ &
$1080^{+119}_{-89}$ \tablenotemark{e} & 1.04 & 31 \\

\ms & 0.831\tablenotemark{h} & 10 57 00.0 & -03 37 36 &$V_{606}\ i_{775}\ z_{850}$ &
1156$^{+82}_{-82}$\tablenotemark{h} & 1.16 &  59 \\   
\cl & 0.834\tablenotemark{i} & 01 52 43.8 & -13 57 19 &$r_{625}\
i_{775}\ z_{850}$ & 919$^{+168}_{-168}$\tablenotemark{j} & 0.92 &  36 \\   
\ctw  & 0.890 & 12 26 58.2 & +33 32 49 &$V_{606}\ I_{814}$  & 1143$^{+162}_{-162}$  & 1.11 & 46 \\                    
\hogb & 0.897\tablenotemark{k} & 16 04 24.0 & +43 04 38 &$V_{606}\ I_{814}$ & 962$^{+141}_{-141}$\tablenotemark{k} & 0.93 & 26 \\    
\hoga & 0.924\tablenotemark{k} & 16 04 33.6 & +43 21 04 &$V_{606}\ I_{814}$  &
640$^{+71}_{-71}$\tablenotemark{k} & 0.61 & 15 \\     
\mei & 1.106\tablenotemark{l} & 09 10 44.9 & 54 22 08.9 & $i_{775}\ z_{850}$ &
675$^{+190}_{-190}$\tablenotemark{l}  & 0.58 & 12 \\   
\jpb & 1.237\tablenotemark{m} & 12 52 48 & -29 27 00 & $i_{775}\ z_{850}$ &
747$^{+74}_{-84}$\tablenotemark{m} & 0.60 & 16 \\
\lynx & 1.260 & 08 48 58.66  & 44 51 57.0 & $i_{775}\ z_{850}$  & 798$^{+208}_{-208}$ & 0.63 & 8 \\
\enddata
\tablenotetext{a}{The number of galaxies classified as early-types
  with $-19.3 > M_B - 1.208 z > -21$.}
\tablenotetext{b}{Imaging area smaller than $2 R_{200}/\pi$}
\tablenotetext{c}{\citet{fisher1998}}
\tablenotetext{d}{\citet{carlberg96}}
\tablenotetext{e}{\citet{halliday2004}}
\tablenotetext{f}{\citet{dressler1999}}
\tablenotetext{g}{\citet{tran2003}}
\tablenotetext{h}{\citet{tran2007}}
\tablenotetext{i}{\citet{blakeslee2005}}
\tablenotetext{j}{\citet{demarco2005a}}
\tablenotetext{k}{\citet{gal2005}}
\tablenotetext{l}{\citet{mei2005a}}
\tablenotetext{m}{\citet{demarco2007}}

\label{summary}
\end{deluxetable*}

\subsection{Morphologies}
\label{morph}

Galaxy morphologies were obtained from the literature for all the
galaxies in our sample.  We use the morphologies from
\citet{dressler1980b} for the $z<0.05$ cluster members.  We removed
all late-type galaxies from that low redshift sample, i.e., all those
classified as spiral, irregular or unknown classifications.  We
considered galaxies classified as S0/a as S0, and galaxies classified
as Sa/0 as spirals.  We list in Table \ref{lowz_num} the number of
early-type galaxies -- E through S0/a -- in the last column.  The
numbers are smaller than the number available in \citet{dressler1980b}
because not all of the galaxies in each cluster are imaged in the SDSS
DR5.

For our higher redshift clusters, we used the classifications of
\citet{dressler97} as tabulated in \citet{smail1997},
\citet{postman2005}, and \citet{desai2007}.  Each cluster was imaged with
either ACS or WFPC2 with morphological classifications done in a
manner consistent with the previous work of \citet{dressler1980b}.
The clusters \ms, \cl, \ctw, \hogb, \hoga, \mei, \jpb, and \lynx\ all
come from \citet{postman2005}.  The MORPHS survey clusters \koo, \od,
\zwcl\ are part of \citet{dressler97} sample.  Five clusters from the
EDisCS survey -- \ecla, \eclb, \eclc, \ecld, and \eclm -- have
morphologies from \citet{desai2007}.  Two additional, \clt\ and \mst,
have morphologies in \citet{fabricant2000} and \citet{tran2003},
respectively.

\section{Data and Measurements}
\label{data}

\begin{figure*}[htbp]
\begin{center}
\includegraphics[width=6.5in]{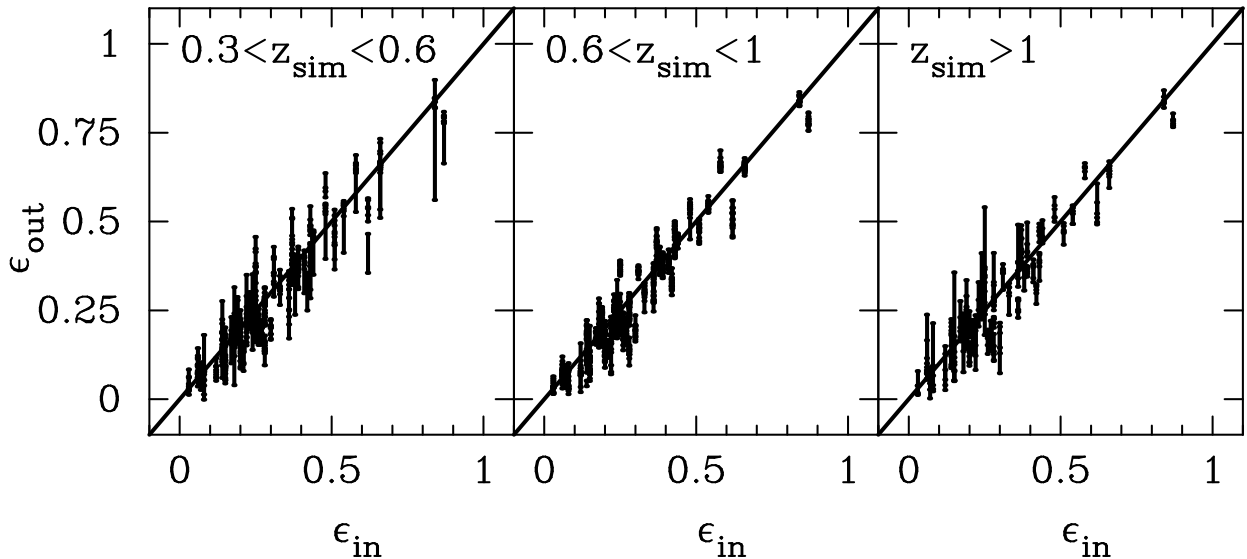}

\end{center}
\caption[f1.eps]{Ellipticity measured from the simulations, which used
  real galaxy images, as a function of the ellipticity in the original
  input image.  The error bars show the scatter around the recovered
  ellipticity values for each galaxy.  Each galaxy image is rescaled
  in size appropriately for the redshift of each cluster.  Each galaxy
  is realized at a variety of signal-to-noise values covering the
  range over which the galaxies in the high redshift sample are
  observed, and were then placed in the cluster imaging data.  We plot
  a straight line with a slope of one, the expected relation if we
  recover the input ellipticities.  The scatter is typically $\sigma_e
  \sim 0.01 - 0.03$.  The scatter increase at lower signal-to-noise,
  and matches the statistical errors from the fitting process.  The
  median offset is $-0.01$ with a range of $0.00$ to $-0.03$, showing
  that our method accurately recovers the ellipticity of these
  low-redshift galaxies when they are observed at high redshift.  No
  systematic trends are seen as a function of ellipticity or
  redshift.}
\label{cl_e}
\end{figure*}

Our sample of early-type galaxies is based on a range in luminosity,
corrected for the observed passive evolution to z $\sim$ 1. The derivation of
the total magnitudes and colors used for our sample selection is
discussed below.  Taking that luminosity-selected sample, we then
discuss the derivation of their ellipticities. These constitute the
key observable for this work, and the implications of those
ellipticity measurements are discussed in the remainder of the
paper. Our final selection will consist of morphologically-selected
early-type galaxies on the red-sequence within a well-defined
magnitude range ($-19.3 > M_B > -21$ at $z= 0$) lying within the
cluster core that is defined by $R_{200}$.  These magnitude limits
correspond to $M^*_B +1 > M_B > M^*_B - 0.75$ and to galaxies with
stellar masses roughly between $10^{10.6}\ M_{\sun} < M < 10^{11.2}\
M_{\sun}$, assuming a ``diet'' Salpeter initial mass function (IMF) 
\citep[see][for a discussion of the IMF and the procedure we use to
estimate the stellar masses of galaxies]{bell2003}.

\subsection{Total Magnitudes and Colors}
\label{measure}

Our samples were selected based on the rest-frame $B$ magnitude.  To
estimate these magnitudes, we used the total magnitude in the passband
closest to the rest-frame $B$.  We also needed a color to correct the
apparent magnitude in the observed passband to a rest-frame $B$
magnitude; how this was done is outlined below. The
S{\'e}rsic model fits that we used to measure the ellipticities also
were used to determine the total magnitudes and the color apertures.
The total magnitude is the normalization of the S{\'e}rsic model fit.
For the color aperture, we used the circularized half-light radius,
$r_{hlr} = a_{hlr} \sqrt q$, where $q$ is the ratio of the minor to
major axis, or $1-\epsilon$, and $a_{hlr}$ is the half-light radius
along the major axis of the best-fitting elliptical model as
determined by GALFIT.  These are the same apertures used in
\citet{mei2008} (see that paper for more detail).

We adjust the magnitudes of the galaxies by $1.208 z$ as measured
for early-type galaxies using the fundamental plane
\citep{vandokkum2006}.  This compensates for the mean passive
evolution of the old stellar population. Our magnitude range covers
$M^{*} - 0.7$ to $M^{*} +1$ using the $M^{*}$ from \citet{norberg2002}
after converting the $b_J$ used by \citet{norberg2002} to the $B$ of
the Johnson-Morgan system \citep{buser1978} that we use in this paper.
We trim our sample at $M_B < -19.3$, as our high redshift samples
become incomplete fainter than that magnitude.  The brighter magnitude
limit is set to be $M_B < -21$ to exclude the most luminous galaxies
(whose formation and evolution may differ, as is explained later in \S
\ref{ellm}).  Thus our adopted magnitude range is  $-19.3 > M_B -
1.208 z > -21$.

Using our simulations of real galaxies that we discuss in more detail
in Appendix \ref{simreal}, we estimate the typical error and offset
for these total magnitude measurements.  We find that total magnitudes
as measured with ACS have an error of $\sigma = 0.10$ mag while
magnitudes measured with WFPC2 have $\sigma = 0.15$ mag.  There is a
small offset at most redshifts, such that we measure a magnitude
brighter than the actual magnitude of the simulated galaxy.  For the
clusters at $0.3 < z < 0.6$, this is only 0.02 mag, while it increases
to 0.06 mag at $0.6 < z < 1$ and 0.10 mag at $z>1$.  This offset is the
same for galaxies regardless of size, morphological type or the value
of $n$ from the S{\'e}rsic model fit.  We do not apply this offset to
the measured magnitudes.  

To ensure a reliable measure of the color, we applied the CLEAN
algorithm \citep{hogbom74} to the
original images in all passbands.  Using CLEAN mitigates the effects
of the different-sized PSF in different passbands,
see \citet{sirianni2005} for examples involving ACS. The final
``CLEANed'' images were used for measuring the galaxy colors within
the $r_{hlr}$ given above.

\subsection{Redshifted Magnitudes}

We transformed the observed magnitudes into redshifted magnitudes
using the same process as \citet{vandokkum96}, \citet{blakeslee2005},
\citet{holden2006} and \citet{holden2007}.  We calculated the
magnitudes of templates in the rest-frame filters.  We then
redshifted the templates, and computed the magnitudes in the observed
filters.  For the templates, we used exponentially decaying
star-formation rate models from \citet[BC03]{bc03}; the same models
were used in \citet{holden2007}.  These models had exponential
time-scales of 0.1- 5 Gyr, covering a range of ages from 0.5 Gyr
to 12 Gyr and three metal abundances, 2.5 solar, 1.0  solar and 0.4 solar.
For the rest-frame filters, we used the $B$ and $V$ curves from
\citet{buser1978}, specifically the B3 curve for the $B_z$ as
tabulated by BC03.  We use the same templates and procedure for all of
the clusters in our sample.

\subsection{Ellipticities}
\label{ellmeas}

We measured the ellipticities for our galaxies using the results from
GALFIT \citep{penggalfit2002}.  GALFIT fits an elliptical model to the
surface brightness profile. Effectively, the ellipticity measurements
we use are ellipticities at the half-light radius. The model is
convolved with a  PSF before it is
compared with the data.  The advantages of the approach we have used
is that the GALFIT fit procedure minimizes
the effect of the smoothing from the PSF.
PSF ``blurring'' will make galaxies appear rounder than they
actually are, unless the galaxies being fitted have sizes much
greater than the PSF.  In Appendix \ref{ellrob}, we discuss in more
detail the robustness of the ellipticity measurements.

This procedure is performed for galaxies at all redshifts, but
requires an estimate of the PSF.  For the clusters of galaxies at
$z<0.05$, our imaging data came SDSS DR5. We use the software tools
provided by the SDSS to extract the PSF appropriate for each galaxy.
For the higher redshift galaxies, the pipeline processing system we
use for ACS (Apsis -- ACS pipeline science investigation software)
provided us with suitable PSF
\citep{blakeslee_pipe2003,blakeslee2005}.  We used empirical PSF
models constructed from multiple ACS observations of 47 Tuc.  For the
WFPC2 data, the PSF still has a strong
positional dependence and is significantly under sampled.  For these
data, we estimated the PSF using the TinyTim software package
\citep{krist95} for each galaxy we fit.  We also used the option in
GALFIT to convolve the model of the galaxy with an over-sampled PSF.
Before comparing the PSF convolved model with the data, GALFIT
rebinned the model to the WFPC pixel scale and smoothed the model with
the charge diffusion kernel.

\subsubsection{Simulations of Ellipticity Measurements}

We constructed a set of simulations using real galaxy images and
placed those images in the data of the high redshift clusters in our
sample.  This is discussed in detail in Appendix \ref{simreal}.  

To summarize, we made multiple measurements of each simulated
individual galaxy to assess both the impact of noise and any
systematic effects, and each galaxy was simulated over a range in
magnitudes.  We found that the scatter in the ellipticity measurements
of the images of real galaxies was $\sigma_e \sim 0.01 - 0.03$ at
magnitudes typical of those in our samples, and only increased (to
$\sigma_e \sim 0.05-0.06$) at or below the magnitude limit of our
samples.  These estimates of the uncertainty are in good agreement
with the errors estimated by the model fitting process used to measure
the ellipticity.  We found no large systematic trends in the
ellipticity measurements.  This can be see in Figure \ref{cl_e} where
we show the recovered ellipticity as a function of the input
ellipticity for three redshift bins. There was only a very small
overall shift of $\delta_e = 0.01$, or 3\% for the typical galaxy,
such that the typical $z>0.3$ galaxy is measured to be slightly
rounder than it would appear in the $z<0.05$ sample.  We found the
shift was the same, regardless of the redshift of the galaxy. No
systematic changes were seen as a function of ellipticity.  For the
rest of the paper, we ignore the small systematic shift of $\delta_e =
0.01$ (noted above and in Figure 1), and simply quote the observed
values.

We also tested the robustness of the ellipticity measurements by using
incorrect PSFs.  We found that this error could result in a
systematic offset of $\delta_e = \pm 0.01$ to $\delta_e = \pm 0.03$
depending on how bad the mismatch between the PSFs were.  The small
size of our statistical errors means that we are sensitive to
systematic errors of this order, and we will discuss the implications
of this in later sections.

\section{Ellipticity Distributions}
\label{elldist}

\subsection{Ellipticity Distributions for the $z<0.05$ Sample}
\label{ellm}
 \begin{figure}[htbp]
 \begin{center}
 \includegraphics[width=3.4in]{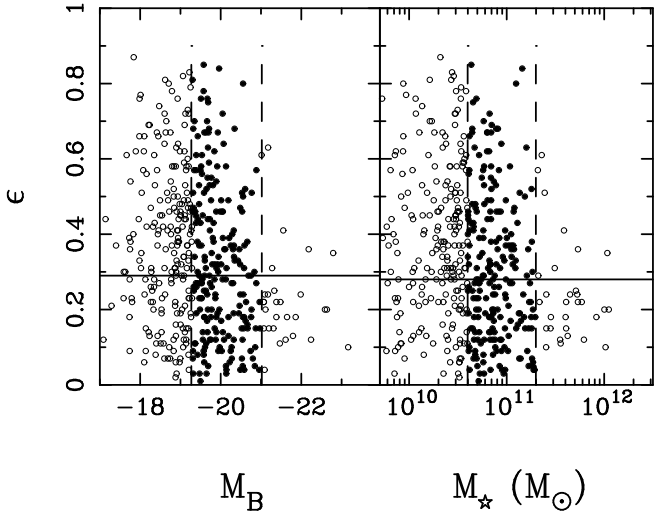}
 \end{center}
 \caption[f2.eps]{Ellipticity versus absolute $B$ magnitude or stellar
mass for the $z<0.05$ clusters, as described in the text in \S
\ref{data}.  The dashed lines show magnitude limits we will use for this
paper, $-19.3 > M_B > -21$.  Below $M_B = -19.3$ our higher redshift
cluster samples become incomplete, while above brighter than $M_B
\sim -21$, the ellipticity distribution of the early-type galaxy
population changes, becoming rounder, possibly the result of the
somewhat different evolutionary history of the most massive galaxies.
The galaxies within our selection limits are filled circles, while all
of the remaining data are shown as open circles. The median
ellipticity of the galaxies in our sample is shown by a solid line.
The mass-selected sample and the luminosity-selected samples yield a
very similar selection with similar median ellipticity, suggesting
that mass-dependent effects are likely to be small.  For the rest of
the paper, we will use the $B$ magnitude limits in the left panel to
select galaxies at higher redshifts.} 
\label{mass_e}
\end{figure}

In this section, we define our selection of $z<0.05$ galaxies, and
discuss how we will characterize the ellipticity distributions in
order to compare them with the $z>0.3$ cluster galaxy sample.

In Figure \ref{mass_e}, we plot the ellipticity as a function of
$M_B$, and also by stellar mass, for all galaxies in the $z<0.05$
sample that are within ${2} R_{200}/{\pi} $, and classified as an
early-type galaxy.  We show the median ellipticity of the whole
population with a solid line. The stellar masses are derived using the
prescription from \citet{bell2003}, which uses the rest-frame colors
to estimate the mass-to-light ratio of the stellar population,
assuming a ``diet'' Salpeter IMF.  The ``diet'' Salpeter IMF is a
Salpeter IMF with truncation at very low masses, resulting in a
mass-to-light ratio of the stellar population 0.15 dex smaller than a
Salpeter IMF.  These stellar mass estimates agree well with the mass
estimates from the fundamental plane \citep[see][]{holden2007}.

At the highest masses or brightest magnitudes, the population becomes
rounder, a result seen in other work \citep[see][for
example]{franx1991,vincent2005}, likely a result of a different
morphological mix among the most luminous galaxies which may have a
somewhat different formation and evolution history.  We will exclude
these most luminous galaxies, $M_B < -21$, from our sample.  We
illustrate our sample range $-19.3 > M_B > -21$ with dashed lines in
Figure \ref{mass_e}.

\begin{figure}[htbp]
 \begin{center}
 \includegraphics[width=3.4in]{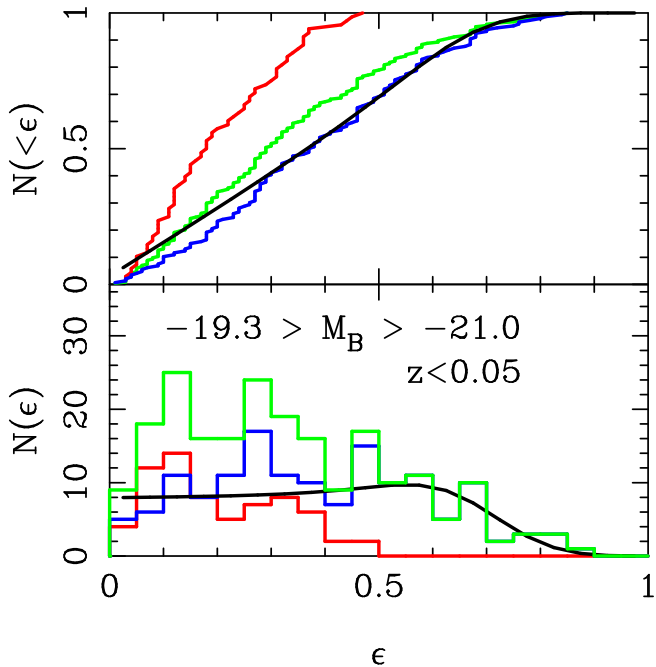}
 \end{center}
 \caption[f3.eps]{Ellipticity distributions for magnitude-selected
   samples by galaxy type for the $z<0.05$ clusters.  We show the
   elliptical and S0 population separately (red - ellipticals, blue -
   S0 galaxies, green - all early-type galaxies).  The two sets of
   galaxies show different ellipticity distributions in our data, as
   expected from the results of previous work.  The elliptical galaxy
   population shows a peaked distribution, with a median ellipticity
   of $\epsilon_{med} = 0.18 \pm 0.01$.  In contrast, the S0 galaxies
   have a broader ellipticity distribution, with $\epsilon_{med} =
   0.38 \pm 0.02$, consistent with a more disk dominated population.
   We show, with a black line, the best-fitting disk population drawn
   from a Gaussian distribution which has a mean thickness of $\bar{b}
   = 0.27 \pm 0.10$ giving a mean ellipticity of $\bar{\epsilon} =
   0.73 \pm 0.10$, with a standard deviation of $\sigma_{\epsilon} =
   0.10 \pm 0.02$.  If the S0 fraction decreases with redshift, the
   overall ellipticity distribution of the cluster population should
   also evolve as there will be fewer galaxies with large
   ellipticities..}
\label{type_e}
\end{figure}

At low redshifts, we know that elliptical and S0 galaxies have
different ellipticity distributions.  In Figure \ref{type_e}, we show
that we reproduce those different distributions with our $z<0.05$
cluster galaxy sample.  The elliptical population, shown in red in
Figure \ref{type_e}, is much rounder than the S0 population, shown in
blue.  The whole of the population is shown in green.  Because S0
galaxies dominate the population, the green line appears closer to the
S0 distribution than the elliptical distribution.  The median
ellipticities are different, with the median ellipticity of S0
galaxies $\epsilon_{med} = 0.38 \pm 0.02$ while the median ellipticity
of elliptical galaxies is $\epsilon_{med} = 0.18 \pm 0.01$.  

In Figure \ref{type_e}, there is a hint of a deficit of round S0
galaxies, a result found by \citet{jf94}.  We fit to the distribution
of S0 ellipticities with the disk galaxy model used in \citet{jf94}.
This disk galaxy distribution is that of an oblate spheroid with the
minor to major axis ratios drawn from a Gaussian distribution with
mean $\bar{\epsilon}$ and a dispersion of $\sigma_{\epsilon}$.  We find
that the values for the Gaussian that best describes the data are
similar to that of \citet{jf94}.  Our values of $\bar{\epsilon} = 0.73
\pm 0.10$ and $\sigma_{\epsilon} = 0.10 \pm 0.02$ are consistent with
$\bar{\epsilon} = 0.65$ and $\sigma_{\epsilon} = 0.10$ from
\citet{jf94} -- a difference $\sim 1 \sigma$.  At small ellipticity
values, the ellipticity distribution of visually classified S0
galaxies fall consistently below the expectation of the model, which
indicates a possible lack of round S0 galaxies.

Coma (A1656) is a dominant contributor to the early-type sample at
$z<0.05$, with about 37\% of the total.  We investigated the impact of
the removal of Coma from the sample.  As can be seen in Table
\ref{coma} doing so makes surprisingly little difference to the
ellipticity measures of the early-types or of the E and S0s
separately.  The ellipticity is consistent within the
uncertainties. The S0 fraction of the $z<0.05$ sample is $68\pm 3$\%
for the whole sample and $72\pm4$\% without Coma.

\begin{deluxetable*}{lrrr}
\tablecolumns{4}
\tablecaption{Median Ellipticities of $z<0.05$ Sample}
\tablehead{\colhead{Sample} & \colhead{All Early-types} & 
\colhead{S0's} & \colhead{Ellipticals} 
  \\
\colhead{} & \colhead{} &  \colhead{}  &  \colhead{}  \\}

\startdata
All $z<0.05$ clusters & 0.29 $\pm$ 0.02 & 0.38 $\pm$ 0.02 & 0.18 $\pm$ 0.01  \\ 
Without Coma & 0.29 $\pm$ 0.02 & 0.34 $\pm$ 0.03 & 0.21 $\pm$ 0.02  \\ 
\enddata
\label{coma}
\end{deluxetable*}

Both \citet{jf94} and \citet{andreon1996} tabulate their
ellipticities.  The median ellipticities in the Coma sample of
\citet{jf94} are $\epsilon_{med} = 0.34 \pm 0.02$ for the S0 galaxies
and $\epsilon_{med} = 0.16 \pm 0.02$ for ellipticals.  Both medians
are slightly rounder than our measured low redshift values, though the
differences are small.  The lower values are possibly due to the use
of the PSF in our analysis.  \citet{jf94} did not remove the smoothing
of the PSF, and so will be offset to slightly rounder values as a
result.

\citet{andreon1996} derive ellipticies for the S0 galaxies and
ellipticals.  These values have been used as a low-redshift comparison
set by other higher redshift studies \citep{smail1997,fasano2000}, it
is useful to understand the difference between these measurements and
ours.  These ellipticities are measured at a fixed $\mu_R$ isophote.
Thus measurement of the ellipticity from \citet{andreon1996} is not
directly comparable to ours and is not the optimal approach for our
study.

\subsection{$z>0.3$ Ellipticity Distributions}
\label{hizelldist}

\begin{figure*}[htbp]
\begin{center}
\includegraphics[width=6.5in]{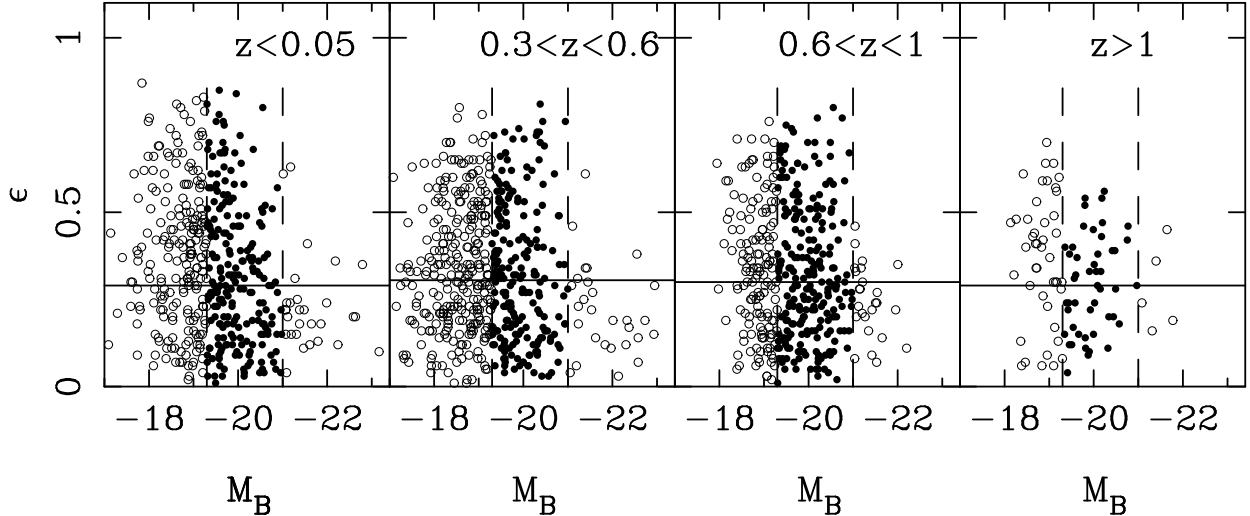}
\end{center}
\caption[f4.eps]{Ellipticity versus absolute $B$ magnitude for all the
  clusters in our sample.  The magnitude range, shown by the dashed
  lines, is the same throughout, as we derived in Figure \ref{mass_e}.
  Each galaxy is selected to be within $2 R_{200} / \pi $ and to have
  $-19.3 > M_B  + 1.208 z > -21$, after removing the effects of passive
  evolution.  We assume early-type galaxies become brighter by
  $1.208 z$ mag \citep{vandokkum2006}.  In each panel, the solid dots
  are those in the magnitude range of our selection, while the open
  circles show the remaining galaxies in our sample.  Each panel
  covers a different range in redshifts, the leftmost, $z<0.05$,
  followed by $0.3 < z < 0.6$, $0.6 < z< 1.0$, and $z>1.0$.  The
  horizontal line is the estimate of the median ellipticity for the
  sample within the magnitude limits. It is striking to note that the
  median ellipticity distribution does not change with redshift.  Only
  at the highest redshifts do we see a hint of fewer highly elliptical
  galaxies.  }
\label{hiz_mass_e}
\end{figure*}

We plot in Figure \ref{hiz_mass_e} the distribution of
ellipticities as a function of absolute $B$ magnitude for four
redshift bins.  As before, we select only those galaxies in a fixed
magnitude range.  We adjust the magnitudes, however, by $1.208 z$ mag to
reflect the amount of passive galaxy evolution that is measured from
the fundamental plane \citep{vandokkum2006}.  We combine each
cluster's sample into three redshift bins, $0.3 < z < 0.6$, $0.6 < z <
1.0$ and $z>1$.  We show, with a solid line, the median ellipticity
for the selected sample in each bin.

\begin{figure}[htbp]
\begin{center}
\includegraphics[width=3.5in]{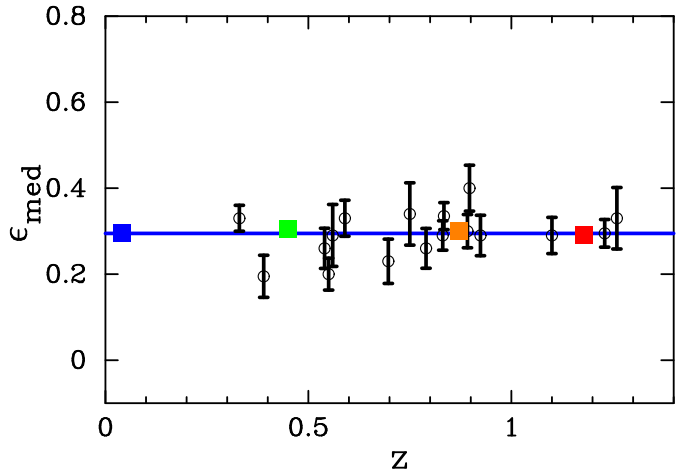}

\end{center}
\caption[f5.eps]{Median ellipticity versus redshift for the clusters
  in our sample.  The open circles are the median ellipticities for
  the early-type galaxies in each cluster, after applying our
  magnitude and radius selection.  The median ellipticity,
  $\epsilon_{med} = 0.29 \pm 0.02$, for all of the galaxies in
  $z<0.05$ cluster sample is shown by the blue square and line.  The
  other squares are the medians for all of the cluster galaxies in the
  redshift ranges $0.3 < z < 0.6$ (green), $0.6 < z < 1.0$ (orange),
  and $z>1$ (red) samples, respectively.  The median ellipticity for
  the whole of the $z>0.3$ sample is $\epsilon_{med} = 0.30 \pm 0.01$.
  The median ellipticity for cluster galaxies in the range of $0.3 < z
  < 0.6$ is $0.31 \pm 0.02$, while the median ellipticity for cluster
  galaxies in the range of $0.6 < z < 1.0$ is $0.30 \pm 0.02$. At
  $z>1$, the median ellipticity is $\epsilon_{med} = 0.29 \pm 0.03$.
  All values are all in excellent agreement with the $0.29 \pm 0.02$
  we find for the $z<0.05$ sample.  We find no individual clusters
  that have drastically different ellipticity distributions.  The lack
  of any trend in the ellipticity, and minimal cluster-to-cluster
  variance, is a striking result. The median ellipticity at $z>0.3$ is
  statistically identical to that at $z<0.05$, being higher by only
  0.01 $\pm$ 0.02.}
\label{avgell}
\end{figure}

We find no evolution in the median ellipticity with redshift. In
Figure \ref{avgell}, we plot the median ellipticity for each cluster.
In this plot we show the median ellipticity for each cluster as open
circles.  For our $z<0.05$ sample, we show just the median value (the
blue square), $\epsilon_{med} = 0.29 \pm 0.02$.  The median values for
the individual $z>0.3$ clusters show larger scatter, but in general
are quite consistent with the low redshift value, with the whole
$z>0.3$ sample having $\epsilon_{med} = 0.30 \pm 0.01$.  The green,
orange, and red points are the median ellipticities (shown also in
Figure \ref{hiz_mass_e} as the horizontal lines) at $0.3 < z < 0.6$
and at $0.6 < z < 1.0$ and $z>1$, $\epsilon_{med} = 0.31 \pm 0.02$,
$\epsilon_{med} = 0.30 \pm 0.02$, $\epsilon_{med} = 0.29 \pm 0.03$,
respectively.  The errors we quote, here and later in the paper, are
the errors on the median, or $\sqrt(\pi/2)$ the error on the mean.  We
confirmed these error estimates with bootstrap resampling.  We
compared the scatter in the high redshift sample by computing the
$\chi^2$ around the low redshift median value.  We find a
$\chi^2_{\nu} = 1.27$ for $\nu=17$ degrees of freedom, confirming both the
good visual agreement between the high redshift data and that the scatter is not higher than expected from random errors.  As we
discuss in Appendix \ref{simreal}, we expect the largest systematic
errors on the median ellipticity values to be $\sim 0.02$, which are not
large enough to shift the high redshift data to significantly smaller
ellipticity values.  The lack of any change in ellipticity with
redshift is striking.  Formally, {\it the median ellipticity of our
  sample of \nmag\ early-type galaxies at $z > 0.3$ is statistically
  identical to that of the \lowzlim\ early-type galaxies at $z <
  0.05$, being higher by only 0.01 $\pm$ 0.02 or 3 $\pm$ 6\%.}

\begin{figure*}[htbp]
\begin{center}
\includegraphics[width=6.5in]{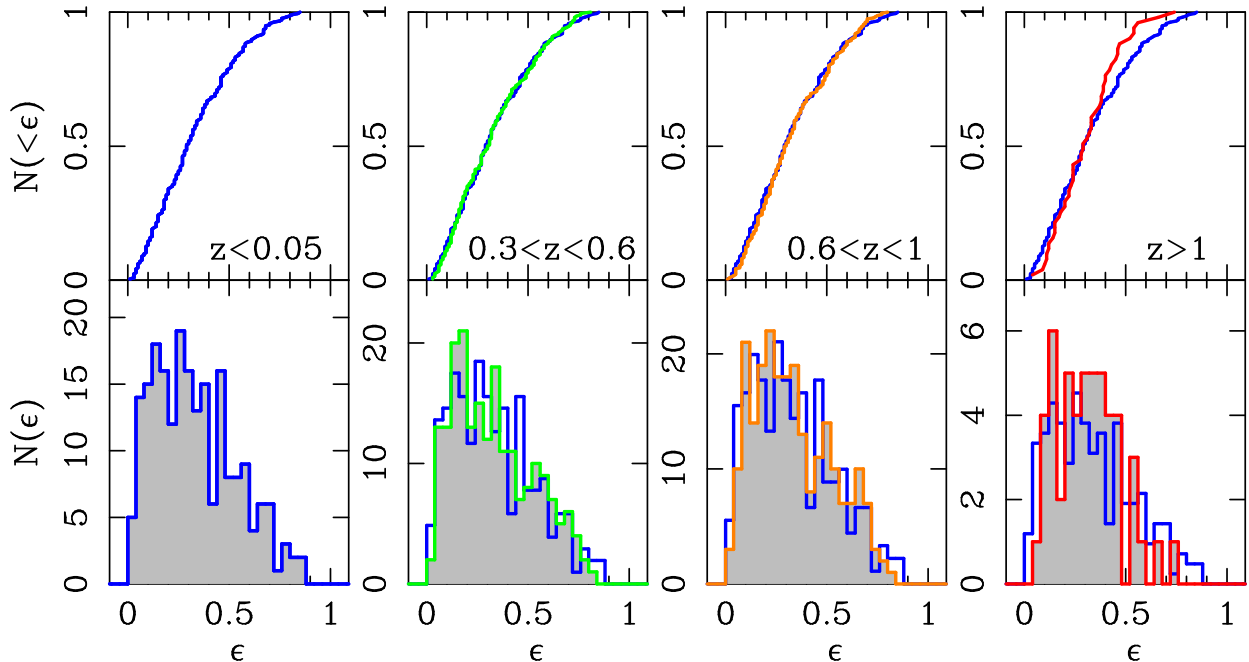}

\end{center} 
\caption[f6.eps]{Cumulative (top) and differential (bottom)
  ellipticity distributions for our four redshift bins.  In each plot,
  we show the distribution of the ellipticities of all early-type
  galaxies in our magnitude and radius selection for all of the
  cluster galaxies in a given redshift bin.  The distributions are
  represented by the shaded histograms, outlined in blue for $z<0.05$,
  green for $0.3 < z < 0.6$, orange for $0.6 < z < 1$, and red for
  $z>1$.  For comparison with the $z>0.3$ cluster samples, we plot our
  sample of $z<0.05$ early-type galaxies in blue, with the
  low-redshift line ``hidden'' when the two lines (frequently)
  overlap.  The $z<0.05$ sample is normalized to have the same number
  of galaxies as each of the $z>0.3$ cluster samples.  The small
  deficit of highly elliptical galaxies at $z>1$ is not statistically
  significant because of the small sample size.  Again, it is striking
  that there is no evolution in the shape of the distribution from
  $z<0.05$ to $z>1$.  {\it The distribution of ellipticities at
    $z>0.3$ agrees with the shape of the $z<0.05$ distribution at the
    1-2\% level (i.e., the probability that they are drawn from the
    same distribution is 98-99\%).}}
\label{sumz_edist} 
\end{figure*}

To add a more detailed assessment of the changes and to provide a more
quantitative basis for the results seen above, we plot the ellipticity
distributions of the samples in Figure \ref{sumz_edist}, both
differentially and in cumulative form.  In each figure, the blue line
shows the $z<0.05$ comparison sample, with the normalization rescaled
to match the higher redshift samples.  At no redshift do we find a
statistically significant change in the ellipticity distribution.  We
use a number of tests to quantify this, including a
Kolmogorov-Smirnov, Wilcoxon-Mann-Whitney rank sum test and a Kuiper
test. Taking the results from these tests we find that {\it the
distribution of ellipticities at $z>0.3$ agrees with the shape of the
$z<0.05$ distribution at the 1-2\% level ( i.e., the probability
that they are drawn from the same distribution is 98-99\%).} This is a
remarkable demonstration of the consistency of the ellipticity
distributions over a time span of more than half the age of the
universe.

There is a hint, in our $z>1$ sample, of a lack of high
ellipticity galaxies, though it is not significant in any of our
tests. This is, in large part, because of the small sample size.
There are only 48 galaxies in the $z>1$ cluster sample. In \S \ref{ellm},
we showed that the median ellipticity for the $z<0.05$ sample was
essentially unchanged when Coma (A1656) was removed (see Table \ref{coma}).
As expected from that test, there is essentially no change in these
results on the ellipticity trends whether Coma is included or removed
from the $z<0.05$ sample.

\section{Discussion}
\label{disc}

We find that neither the median ellipticity nor the shape of the high
redshift ellipticity distribution of early-type cluster galaxies
evolves with redshift, implying no change in the overall distribution
of the bulge-to-disk ratio of early-type galaxies with redshift.  As
we show in Figure \ref{type_e}, the S0 population at low redshift
($z<0.05$) has a different ellipticity distribution than the
elliptical population, with S0 galaxies having a higher median
ellipticity.  If the S0 fraction of the early-type galaxy population
decreases, then the median ellipticity of the early-type population
should decrease.  The lack of evolution we observe in the median
ellipticity and in the shape of the ellipticity distribution implies
little or no evolution in the S0 fraction.  This differs from the
expectation from previous work, such as that of \citet{dressler97},
which finds a decrease in the S0 fraction with redshift.

\subsection{Morphological Evolution in the $z>0.3$ Cluster Sample}
\label{morphevol}

\begin{figure}[htbp]
\begin{center}
\includegraphics[width=3.4in]{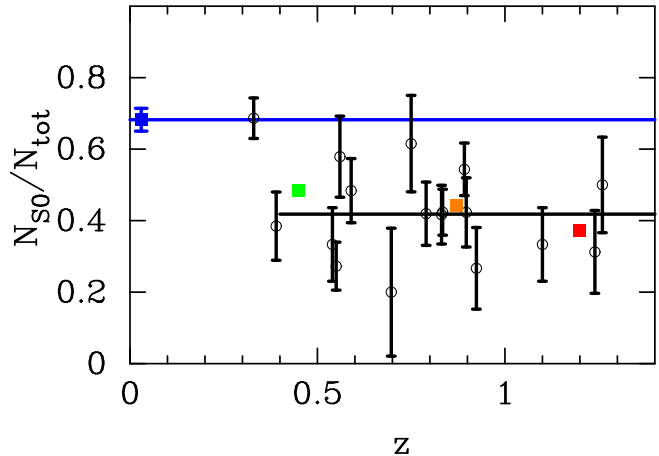}

\end{center} \caption[f7.eps]{Ratio of morphologically identified
S0 galaxies from the studies discussed in \S \ref{morph} to the total
number of elliptical and S0 galaxies.  The fraction of S0 cluster
galaxies at $z>0.4$ is lower than that seen at $z<0.05$, as seen in
previous work.  We compute the average fraction of S0 galaxies in the
early-type population for all of the $z>0.3$ clusters and plot that
value, $42 \pm 2$\% as a solid black line.  For contrast, we show the
$z<0.05$ value, $68 \pm 3$\% , as a solid blue line.  We plot, as
squares, the average values of the S0 fraction of early-type galaxies
in three redshift bins, $0.3 < z < 0.6$ (green), $0.6 < z < 1.0$
(orange) and $z >1$ (red).   Our sample shows the same trend in S0
fraction found by other work. However, the very different fraction of
galaxies classified as S0's at $z>0.4$ is in seeming contradiction
with the lack of evolution in the ellipticity distribution.}
\label{s0frac}
\end{figure}

One possible reason why we may not find any evolution in the
ellipticity distribution of the cluster early-type population could be
because of the nature of our sample.  The fraction of galaxies
morphologically selected as S0 galaxies may not change in our sample
as has been found by other authors.  However, when we look at the S0
fraction of our sample with redshift, using the visual classifications
for our sample from the literature, we find a similar trend to what
has been reported in other papers, i.e., a lower fraction of S0
galaxies at redshifts $z>0.4$.  In Figure \ref{s0frac}, we plot the
fraction of morphologically identified S0 galaxies with redshift (from
the studies discussed in \S \ref{morph}).  Our sample has properties
consistent with previous work
\citep{dressler97,fasano2000,postman2005,desai2007} as would be
expected since our sample largely overlaps with previous studies and
uses visual classifications from those studies.  We find 42 $\pm$ 2\%
of the early-type galaxy population are S0 galaxies at $z>0.4$. For
comparison, the fraction of S0 galaxies within $2 R_{200}/\pi$ for the
\citet{postman2005} ``$z\sim1$ composite'' sample is 35 $\pm$ 3\%.
This shows that our red-sequence selection is consistent, being less
than 2$\sigma$ different (the Postman \etal\ 2005 value is derived by
summing the individual listings in that paper's Table 4 under the item
labeled ``$z\sim1$ composite'').

Our $z<0.05$ sample of early-type cluster galaxies, selected in the
same manner as our $z>0.3$ sample, has a S0 fraction of $68 \pm 3$\%.
This fraction is consistent with other analyses.  Examining the whole
sample of early-type galaxies within $2 R_{200}/\pi$, we find a S0
fraction of $67 \pm 3$\% for all of the early-type galaxies within for
the 10 clusters we use from \citet{dressler1980b}, in good agreement
with that found by \citet{dressler97}.  Therefore, our red-sequence
selection produces a comparable sample of early-type cluster galaxies
to those from previous efforts.

We find no statistically significant evidence for evolution within the
$z>0.4$ cluster sample, which is also consistent with most of the
previous work \citep[e.g.,][]{fasano2000,postman2005,desai2007}.
Using the visual classifications, our sample also shows that evolution
in the S0 fraction occurs between $z\sim 0$ and $z\sim 0.4$ as the
authors above have noted.  Note that the fractions we are discussing
here are the fraction of S0 galaxies in the early-type galaxy
population \citep[see][for a discussion of the size of the systematic
error in separating the elliptical and S0 populations.]{postman2005}

The result from our measurements of the ellipticity distributions in
our clusters is that there is essentially no evolution in the
ellipticity distributions from $z\sim1$ to $z\sim 0$, and thus that
there is no change in the overall distribution of the bulge-to-disk
ratio of early-type galaxies.  If we assume no evolution in the
bulge-to-disk ratio distributions of the elliptical and S0 population
separately as others have done, we can conclude that the E/S0 ratio
does not evolve over this interval.  This contrasts with the clear and
significant evolution in the S0 fraction seen from many studies
\citep[e.g.,][]{dressler97,fasano2000,postman2005,desai2007} based on
visual classifications. While one explanation is that there is a
problem with the visual classifications, this raises an interesting
issue.  In \citet{postman2005} and \citet{desai2007}, what evolves at
$z>0.4$ is the fraction of S0 galaxies and the fraction of spiral and
irregular galaxies.  In contrast, the fraction of elliptical galaxies
does not evolve.  If the misclassification of round S0 galaxies as
ellipticals alone causes the fraction of S0 galaxies to decrease, it
is puzzling that there is no corresponding increase in the fraction of
elliptical galaxies in \citet{postman2005} and \citet{desai2007}.
Rather, in previous work, the spiral fraction increases, and there is
little expectation that face-on S0 galaxies would be misclassified as
spirals.  The source of this difference remains to be resolved.  Our
study does not provide an opportunity for a resolution of this issue,
but we wanted to highlight it for further work by others.

\subsection{Ellipticity by Galaxy Type}

\begin{figure}[htbp]
\begin{center}
\includegraphics[width=3.5in]{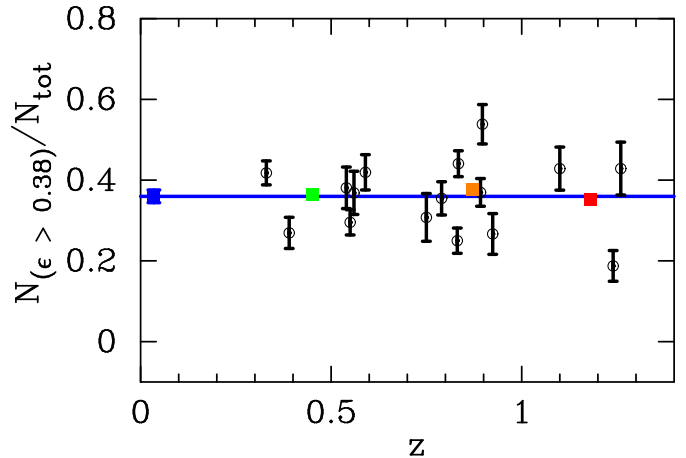}

\end{center}
\caption[f8.eps]{Fraction of cluster early-type galaxies with
  ellipticities greater than the median ellipticity of the $z<0.05$ S0
  galaxies.  The open circles are the fractions of galaxies with
  $\epsilon_{med} > 0.38$, in each cluster.  The squares are the
  fractions for the $z< 0.05$ (blue), $0.3 < z< 0.6$ (green), $0.6 < z
  < 1.0$ (orange), and $z>1$ (red) samples respectively.  At low
  redshift, this fraction has only a small contamination from
  elliptical galaxies, such that the fraction of galaxies with
  $\epsilon_{med} > 0.38$ is the half fraction of S0 galaxies with a
  5\% contamination from elliptical galaxies.  If the observed lack of
  evolution in the median ellipticity of the $z>0.3$ population is a
  result of evolution in the ellipticities of elliptical galaxies
  masking the decline in the S0 population, we should still see a
  change in the population of galaxies with $\epsilon_{med} > 0.38$.
  Galaxies with $\epsilon_{med} > 0.38$ represent the most inclined of
  the disk-dominated early-type galaxies at low redshift.  The lack of
  evolution in this fraction suggests little evolution in the fraction
  of disk-dominated early-type galaxies as a whole.}
\label{sharp}
\end{figure}

Our simulations show that the estimates we use for ellipticities are
robust, and do not show a large systematic change with redshift (see
Figure \ref{cl_e}.)  Some of the observed change in the fraction of S0
galaxies in Figure \ref{s0frac} may come about from a
misclassification of round galaxies as ellipticals, as others
\citep{jf94} have suggested and as we have discussed earlier.  It
would be useful to examine a sample of S0s for which the elliptical
contamination was, likely, very small. We examined the fraction of
galaxies with ellipticities above the median of the $z<0.05$ S0
population, $\epsilon_{med} = 0.38$.  Measuring the fraction of
galaxies with $\epsilon_{med} > 0.38$ should give us an estimate of
the S0 population that we expect to be modestly contaminated by
elliptical galaxies.  In our $z<0.05$ sample, we find that $5 \pm 3$\%
of galaxies with $\epsilon > 0.38$ are ellipticals.  We plot the
fraction of galaxies with $\epsilon_{med} > 0.38$ in Figure
\ref{sharp} as a function of redshift.  This once again, shows no
change with redshift, implying a S0 fraction that does not evolve
under the assumption that the ellipticity distribution of ellipticals
and S0 galaxies do not separately evolve.

\begin{figure}[htbp]
\begin{center}
\includegraphics[width=3.4in]{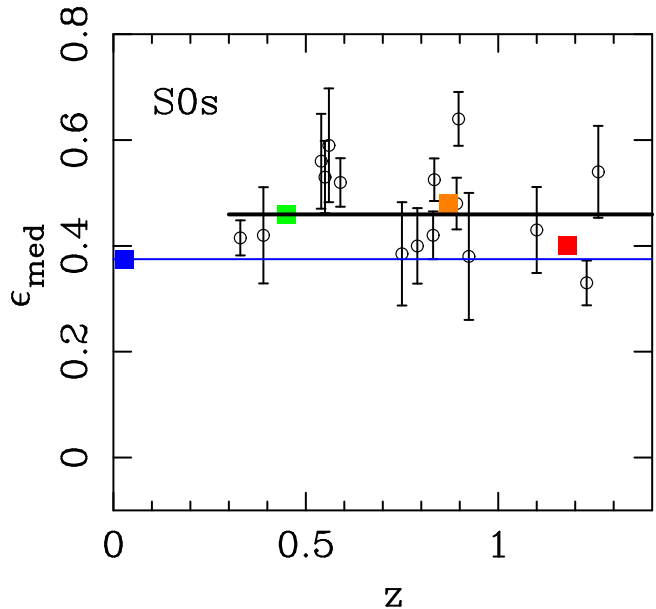}

\end{center}
\caption[f9.eps]{Median ellipticity versus redshift for cluster S0
  galaxies.  The open circles are the median ellipticities for the
  galaxies in each cluster.  The blue squares are the median for all
  of the galaxies in the $z<0.05$ cluster sample.  The other squares
  are the medians for the $0.3 < z< 0.6$ (green), $0.6 < z < 1.0$
  (orange), and $z>1$ (red) samples, respectively.  We also show the
  median value for the $z<0.05$ S0 sample as a blue line.  We show,
  with a black line, the median ellipticities of all the S0 galaxies
  in our $z>0.3$ sample. The median ellipticity of the S0 population
  increases at $z>0.3$, with $\epsilon_{med} = 0.47 \pm 0.02$.  The
  median ellipticity of the $z<0.05$ S0 galaxy is $\epsilon_{med} =
  0.38 \pm 0.02$.  Such a change in the ellipticity of galaxies
  classified as S0s can explain how the fraction of galaxies
  classified as S0's decreases while, at the same time, the median
  ellipticity of the cluster galaxy population as a whole stays the
  same.}
\label{avge_type}
\end{figure}

If the S0 fraction decreases with redshift while the median
ellipticity of the early-type population stays the same, we expect that
the ellipticity distribution for the S0 galaxies should increase.  In
Figure \ref{avge_type}, we plot the median ellipticity of galaxies
that are classified visually as S0 galaxies.  We show that the median
ellipticity of the S0 population is higher in the higher redshift
clusters.  In the $z > 0.3$ sample, we find that the median
ellipticity of the S0 galaxy population is $\epsilon_{med} = 0.47 \pm
0.02$ as compared with $\epsilon_{med} = 0.38 \pm 0.02$ in our $z <
0.05$ low redshift sample\footnote{These median values are the medians
  of the all S0 galaxies that meet our selection criteria. These are
  not the median of the data points we plot in Figure \ref{avge_type}.
  The error we quote is the error on the median, or $\sqrt(\pi/2)$ the
  error on the mean.  We confirm these errors with bootstrapping which
  yields good agreement.}. The difference between the $z > 0.3$ and
the $z < 0.05$ S0 samples medians is significant at the $>3\sigma$
when using a Student's t-test.  Restricting the redshift range to
$z>0.4$, the redshift where other authors have reported that the S0
fraction significant evolves, we find $\epsilon_{med} = 0.48 \pm
0.02$, which is also a $>3\sigma$ difference.  The significance
remains $>3\sigma$ regardless of whether Coma is included in the
$z<0.05$ sample or not. 

\begin{figure*}[htbp]
\begin{center}
\includegraphics[width=6.5in]{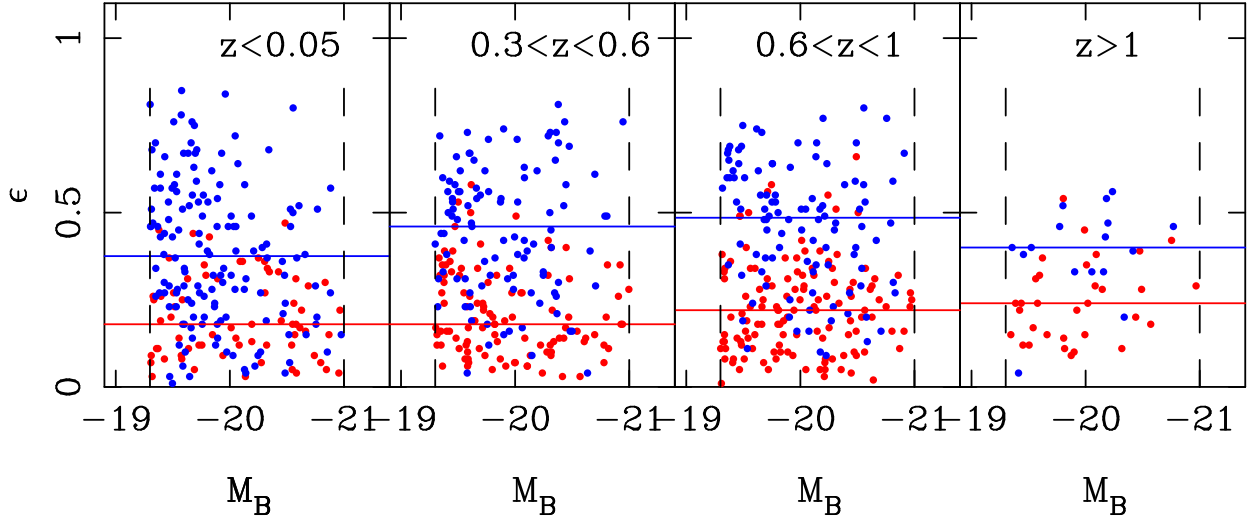}

\end{center}

\caption[f10.eps]{The distribution of ellipticity versus $M_B$ for
  cluster elliptical (red) and S0 (blue) galaxies in four redshift
  bins. The solid points are our sample within our magnitude limits of
  $-19.3 > M_B + 1.208 z > -21$.  This plot is similar to Fig.\
  \ref{hiz_mass_e}, except that we color code the S0 and elliptical
  galaxies separately.  We show the median values of the populations
  as lines, red for the ellipticals and blue for the S0 galaxies.  We
  find no statistically significant evolution in the elliptical
  population.  There is a statistically significant ($>3\sigma$)
  change in the distribution of ellipticities for S0 galaxies when
  comparing the $z<0.05$ S0 population with $\epsilon_{med} = 0.38 \pm
  0.02$ and the $0.3<z<0.6$ ($\epsilon_{med} = 0.46 \pm 0.02$) or
  $0.6<z<1$ sample ($\epsilon_{med} = 0.49 \pm 0.02$).  The
  uncertainty in the highest redshift sample is too large for any
  useful comparisons to be made.  The change in the ellipticity of the
  elliptical galaxies with redshift is not statistically significant.
  From this figure we conclude that the increase in the median
  ellipticity seen in Fig.\ \ref{avge_type} could arise from a lack of
  round S0 galaxies at redshifts higher than $z>0.3$. }

\label{hiz_mass_etype}
\end{figure*}

In Figure \ref{hiz_mass_etype} (similar to Figure \ref{hiz_mass_e},
but with just the complete sample with E and S0 identified), we plot
the ellipticities as a function of absolute magnitude in four redshift
bins. It appears from this figure that there is a deficit of rounder
S0 galaxies (blue points) at higher redshifts (excluding the $z>1$ bin
which is less complete and has poor statistics), implying that the S0
galaxies at $z > 0.3$ are drawn from a different distribution than
those at $z < 0.05$. The progression of higher ellipticities for the
S0 population thus appears to come about from a change in the shape of
the ellipticity distribution of S0 galaxies at higher redshift as
compared with those at lower redshifts.  We discuss this in \S \ref{evols0}.

\begin{figure*}[htbp]
\begin{center}
\includegraphics[width=6.5in]{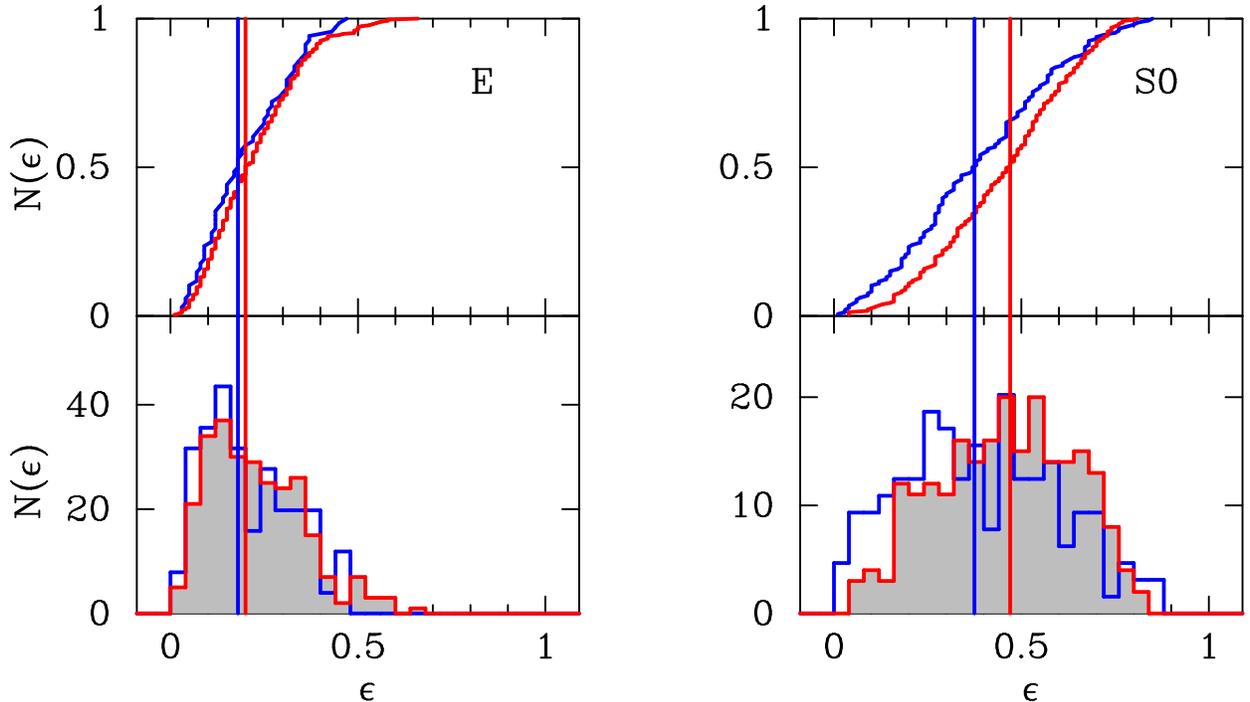}

\end{center}
\caption[f11.eps]{Ellipticity distributions for cluster elliptical
  (left) and S0 (right) galaxies.  The $z<0.05$ sample is shown in
  blue, while the $z>0.3$ sample is shown in red, with the median
  ellipticity for each shown as a appropriately colored vertical line.
  We find that the $z>0.3$ elliptical population is statistically the
  same as the $z<0.05$ population.  In contrast, there is a
  statistically significant difference ($>3\sigma$) in the
  ellipticities of the S0 population.  This appears to be a deficit of
  round S0 galaxies.  We find 56 S0 galaxies with $\epsilon < 0.3$ in
  our $z>0.3$ sample, whereas we expect 95 S0 galaxies,
  after rescaling the $z<0.05$ to match the total number of expected
  S0 galaxies.  This reinforces Fig.\ \ref{avge_type}, showing that the
  typical ellipticity for S0 galaxies at $z>0.3$ is higher than at
  $z<0.05$ and it appears that there is a deficit of round S0 galaxies
  which is causing this difference in the ellipticity distributions.
}
\label{hizell_type_dist}
\end{figure*}

In Figure \ref{hizell_type_dist}, we quantify what we see in Figure
\ref{hiz_mass_etype}. We plot the distributions of the ellipticities
for the $z > 0.3$ elliptical and S0 galaxies separately.  We find that
the shape of the $z>0.3$ S0 distribution differs from the
$z<0.05$ distribution, in contrast with the elliptical distributions
which show little change between the two samples.  For example, if we
examine the number of S0 galaxies with $\epsilon < 0.3$, we find that
there are 56 in the $z>0.3$ sample, whereas we expect 95 from the
scaled low redshift sample of S0 galaxies.  We compared the shapes of
the two S0 ellipticity distributions using a number of statistical
tests. The probability of such a change in the ellipticity
distribution of S0 galaxies is $>3\sigma$ when using a
Kolmogorov-Smirnov, a Wilcoxon-Mann-Whitney rank sum test, and a Kuiper
test. These are the same tests that showed no evolution in the
ellipticity distribution in \S \ref{hizelldist}.  These tests
reinforce the result, from the analysis of the trends in Figure
\ref{avge_type}, that the cluster S0 galaxies have a different
ellipticity distribution at $z > 0.3$ than at $z < 0.05$.  These
trends, both in the median and in the shape of the distribution, are
present regardless of how we select the higher redshift sample,
whether $z>0.3$, $z>0.4$ or $0.4 < z < 1$.  In all cases, the
distributions differ from the $z<0.05$ sample at $>3\sigma$.

Our work is not the only one that finds a different median ellipticity
for S0 galaxies at higher redshift in cluster samples.
\citet{moran2007} give the morphological classifications and
ellipticities of a sample of S0 galaxies at $z\sim 0.45$.  The
authors measured the ellipticities of the cluster galaxies using the
same approach and software that we use (and thus their ellipticities
are PSF-corrected.)  For their cluster S0 galaxies, \citet{moran2007}
find $\epsilon_{med} = 0.48 \pm 0.04$, in excellent agreement with our
value of $\epsilon_{med} = 0.48 \pm 0.02$.

Previous work found little or no evolution in the ellipticities of
S0 and elliptical galaxies.  However, most of the previous work relied
on measuring the ellipticities of galaxies using flux-weighted
moments of galaxies, without taking into account the ``blurring'' effects
of the PSF or the lower resolution of observations
at higher redshifts.  We show, in Appendix \ref{ellrob}, that the
ellipticity measurements of galaxies based on simple
flux-weighted moments are biased quite significantly towards
rounder values for the ACS and WFPC2 data that has been used without
correcting for the effects of smoothing by the PSF.

\subsection{Constraints on the Evolution of the Bulge-to-Disk Ratio}
\label{btod}

The observed lack of any evolution in the ellipticity distribution for
early-type galaxies has interesting implications for the distribution
of the bulge-to-disk ratios of the early-type population from $z \sim
0$ to $z \sim 1$.  The average observed ellipticity is directly
related to the average intrinsic ellipticity \citep{binneym1998}.  If
the average bulge-to-disk of a population of galaxies changes, the
average intrinsic ellipticity must change too, and vice versa.  Our
measurement of an unchanging ellipticity distribution implies that
the average bulge-to-disk remains constant.

We performed simple simulations to derive the average ellipticities
using models of galaxies with a variety of bulge-to-disk ratios
projected over all possible viewing angles with a fixed scale length
for the bulges and disk.  We find a roughly linear relation between
the average bulge-to-total ratio and the average ellipticity of the
population.  For example, for these simple models, a change in the
average bulge-to-total ratio from 0.1 to 0.3, or from 0.3 to 0.6,
would be observed as a shift of $\delta \bar{\epsilon} = 0.08$ in the
average ellipticity.  Such a shift would be a $4\sigma$ change for our
samples (see Figure \ref{avgell}.)  While one could potentially
develop a model whereby the underlying scale lengths relatively evolve
with redshift, this seems rather contrived.  ``Occam's Razor'' would
lead one to prefer a model where a constant ellipticity distribution
implies a constant distribution of bulge-to-disk ratios over the
redshift range from $z\sim1$ to $z\sim0$.

\subsection{Implications for Evolution in the Fraction of Disk-Dominated Early-type Galaxies}
\label{evols0}

We performed simulations to see if the observed S0 fraction of 42\%,
for the sample of clusters with $z>0.4$, is consistent with the lack
of evolution we see in the ellipticity distribution.  We constructed
10,000 mock catalogs for the clusters in the $z>0.4$ sample.  For each
cluster, we assumed that the fraction of S0 galaxies was 42\%.  We
randomly drew the ellipticities for the S0s from the observed
distribution of S0s at $z<0.05$, and drew the ellipticities for the Es
from the observed low redshift distribution for Es (see Figure
\ref{type_e}.)  We added the uncertainties and systematics applicable
at high redshift as detailed in \S \ref{measure} and Appendix
\ref{simreal} The resulting ellipticity distributions had median
ellipticities lower than the observed value (0.29) for all 10,000 of
the mock catalogs, with a typical value of $\epsilon_{med}=0.25$.
This rules out at $>4\sigma$ the assumption that the underlying
ellipticities and S0 fraction are both constant with redshift.

Our others tests which we use to compare the $z<0.05$ sample and the
$z>0.3$ sample give similar results (the Kolmogorov-Smirnov statistic,
a Wilcoxon-Mann-Whitney rank sum statistic and a Kuiper
statistic). None of the mock catalogs show the same good agreement
between the $z>0.3$ sample and $z<0.05$ sample in the ellipticity
distribution as we observe.

We repeated these simulations for a variety of S0 fractions.  We find
that the S0 fraction value that best matches the $z>0.4$ ellipticity
distribution is 66\% with a $1\sigma$, or 68\% confidence limits, of
$\pm$6.5\%.  Another way to consider this is a $1 \sigma$ change in
the median ellipticity at $z>0.3$, a change of
$\delta \epsilon=0.02$, would imply a change in the S0 fraction of
6.5\%.  We can rule out, at the 95\% confidence limit, S0 fractions
below 53\%, and we can rule out the observed fraction of visually
classified S0 galaxies of 42 $\pm$ 2\% at the $>99.9\%$ confidence
limit or at the 4$\sigma$ level.

To reconcile the evolution seen in the S0 fraction in the
visually-classified samples with the lack of evolution in the
ellipticity distribution requires that either the S0 and elliptical
populations evolve in ellipticity in such a way that the combined
samples shows no evolution, or that some fraction of the S0 population
has been misclassified as other morphological types
\citep[cf.,][]{jf94,blakeslee2005}.

\subsubsection{Possible Spiral Contamination?}

One possibility is that the higher ellipticity, or more edge-on, S0
population is contaminated with misclassified red spirals.  This would
cause the S0 fraction to be overestimated, while, at the same time
change the shape of the ellipticity distribution.  The spiral fraction
among galaxies in the mass range of the cluster sample in our work is
$\sim10-15$\% \citep{holden2007} while the fraction of dusty, red
objects is $\sim10$\% \citep{saintonge2008}.  Since the
misclassification would be predominately for edge-on systems, these
fractions are likely upper limits for the fraction of spirals
misclassified as S0 galaxies.  Is it possible that such a small fraction
of ``spiral'' galaxies could  contaminate the population
visually classified as S0 galaxies and change the ellipticity
distribution at higher redshift?  We examined the ellipticities of the
galaxies classified as spirals in both the $z<0.05$ and $z>0.3$ sample
to see if there is a deficit of edge-on spirals.  We found the
opposite, that there is a shift in the high redshift spiral population
to larger ellipticities, $\epsilon_{med} = 0.38 \pm 0.04$ at $z<0.05$
\citep[as expected, see][]{ryden2006} to $\epsilon_{med} = 0.45\pm
0.03$ at $z>0.3$, but it is not statistically significant.  From this
we conclude that a red spiral population is not a major source of
contamination, especially at large ellipticities.

Given the quantitative nature of the current study we feel that it
provides a key datum for consideration of the evolutionary history of
early-type galaxies over the last 8 Gyr.  The result that the
ellipticity distribution is essentially unchanged from $z\sim 0$ to
$z\sim 1$, combined with the result that the fraction of highly
elongated disk-dominated systems does not evolve (see Figure
\ref{sharp}), provides strong evidence that the overall
bulge-to-disk ratio distribution of the population does not evolve.

\section{Summary and Implications}
\label{conclusion}

We have compiled a sample of 10 $z>0.05$ clusters of galaxies and a
comparable sample of 17 $z>0.3$ clusters of galaxies with {\em HST} imaging.
For each cluster, we selected a subsample of galaxies that lie on the
red sequence and have been classified as an early-type galaxy
(elliptical or S0 galaxy).  To derive a robust sample at all
redshifts, we selected galaxies to fall within a magnitude range of
$-19.3 > M_B+ 1.208 z > -21$ and required the galaxies to lie within
$2 R_{200}/\pi $ (to ensure that the de-projected sample lies
within $R_{200}$) of the cluster center.  We change the
rest-frame $B$ magnitudes by $1.208 z$ to match the mass-to-light
evolution of cluster early-type galaxies as measured by the
fundamental plane \citep{vandokkum2006}.  The magnitude limits
correspond to $M^*_B +1 > M_B > M^*_B - 0.75$ or galaxies with stellar
masses roughly between $10^{10.6}\ M_{\sun} < M < 10^{11.2}\
M_{\sun}$, assuming a ``diet'' Salpeter IMF.

We performed extensive simulations of the ellipticity measurements
to test their robustness and to minimize systematics with redshift.
We found that it is crucial to use model fits (e.g., GALFIT)
where the effect of the PSF can be modeled
and removed.  From our extensive simulations we find that we can
robustly measure the ellipticity of cluster early-type galaxies (by
using PSF-corrected model fits) out to $z\sim 1.3$ at magnitudes
that allow the galaxies to also be morphologically classified.  The
systematic error we find in our simulations is only $\delta_{\epsilon}
= -0.01$, or a 3\% change for the median galaxy.  We do not apply this
small correction to our data.  We find that not including the effect
of the PSF can cause systematic shifts of
$\delta_{\epsilon} = -0.10$, or a 30\% change for the median
galaxy.

Using our two samples of cluster early-type galaxies, we measure the
evolution in the ellipticity of $z>0.3$ cluster early-type galaxies 
compared with a substantial sample at $z<0.05$.

\begin{enumerate}

\item We find no evolution in the median ellipticity of $z>0.3$
  cluster early-type (elliptical and S0) galaxies relative to the low
  redshift sample. The median ellipticity at $z<0.05$ is
  $\epsilon_{med} = 0.29 \pm 0.02$, and is $\epsilon_{med} = 0.30 \pm
  0.01$ at $z>0.3$.  {\it The median ellipticity of our sample of 487
    early-type galaxies at $z > 0.3$ is statistically the same as that
    of the 210 early-type galaxies at $z < 0.05$, being higher by only
    0.01 $\pm$ 0.02 or 3 $\pm$ 6\%.}

\item The shape of the ellipticity distribution of $z>0.3$ galaxies
  also does not evolve. {\it The distribution of ellipticities at
    $z>0.3$ agrees with the shape of the $z<0.05$ distribution at the
    1-2\% level (i.e., the probability that they are drawn from the
    same distribution is 98-99\%).}

\item Using visual classifications from previous reference studies, we
  find that our sample shows a similar decrease in the fraction of
  early-type galaxies classified as disk-dominated systems (S0s) at
  $z>0.4$.  As other studies have found, t he S0 fraction is 68$\pm$3\%
  at $z<0.05$ and decreases to 42$\pm$2\% at $z>0.4$.

\item For the S0 fraction to decrease with increasing redshift, while
  the median ellipticity of the overall cluster early-type population
  stays the same, the median S0 ellipticity needs to be larger at
  higher redshift. There is a trend in our data for $z>0.3$ in this
  sense.  The median S0 galaxy at $z<0.05$ is $\epsilon_{med} = 0.38
  \pm 0.02$ while at $z > 0.3$, we find a $\sim 3\sigma$ different
  $\epsilon_{med} =0.47 \pm 0.02$ for galaxies classified as S0s.
  Whether this change is large enough to account for the change seen
  in the morphological sample remains to be determined.  We explored
  if the change in the ellipticity distribution of S0 galaxies could
  come from a deficit of round S0 galaxies or from the addition of
  misclassified spiral galaxies.  However, we found that neither of
  these simple scenarios can explain our results.

\end{enumerate}

Our results on the unchanging ellipticity distributions lead us to
conclude that there has been little or no evolution in the overall
distribution of bulge-to-disk ratio of early-type galaxies over the redshift
range $0 < z < 1$ for morphologically-selected samples of early-type
red-sequence galaxies with $M^*_B + 1 > M_B > M^*_B - 0.75$ in the
dense cores of clusters inside $R_{200}$. In particular, our results
allow us to rule out a S0 fraction of $<$47-51\% at $z>0.3$ at the
3$\sigma$ level assuming no evolution in the ellipticity distributions
of elliptical and S0 galaxies. {\it If we assume, as in all previous
  studies, that the intrinsic ellipticity distribution of both
  elliptical and S0 galaxies remains constant, then we conclude from
  the lack of evolution in the observed early-type ellipticity
  distribution that the relative fractions of ellipticals and S0s do
  not evolve over the last $\sim8$Gyr, or from $z\sim1$ to $z=0$,
  for a red-sequence selected sample of early-type galaxies in the
  cores of clusters of galaxies.}

These results do highlight an inconsistency within the wide range of
studies that have occurred on the evolution of early-type galaxies
over the last decade.  Over this redshift range, and particularly
since $z\sim0.4$, the fraction of morphologically-identified S0
galaxies from visual classifications has been found to be
significantly lower than the $z\sim0$ value \citep[see, e.g.,
][]{dressler97,fasano2000,postman2005,desai2007}. Reconciling the
trends seen in the elliptical, S0 and spiral fractions, as discussed
in \S \ref{morphevol} remains to be understood.                

The early-type galaxy population, both in the field and in clusters,
does not evolve purely passively. The volume-averaged number density
of red galaxies, which are mostly early-type galaxies
\citep{bell2004b}, grows by a factor $\sim2$ between $z\sim 1$ and the
present \citep{bell2004,brown2007,faber2007}.  For the cluster
population such evolution is harder to quantify, but mergers
\citep{vandokkum1999,tran2005} and filaments around massive clusters
at $z\sim 0.8$ \citep[e.g.,][]{kodama2005,patel2008} suggest that
interactions and infalling galaxies enhance and modify over time the
galaxy population in the cluster core. Besides evolution in the
population of early-type galaxies, individual early-type galaxies also
change over time. As pointed out by, e.g., \citet{jorgensen2005}, 
the evolution of the line strengths is not compatible with purely
passive evolution.  Moreover, early-type galaxies at $z\sim1$
were recently demonstrated to be significantly smaller than today by a
factor of two \citet{vanderwel2008c}. Hence, the resulting picture is
complicated: early-type galaxies, both as individual objects and as a
population, undergo substantial changes between $z\sim 1$ and the present.

On the other hand, many basic properties of the early-type population
have remained unchanged over the past $\sim8$Gyrs, providing useful
means for characterizing and quantifying the observed evolution. At
any redshift $z\sim1$ early-type galaxies occupy a tight
color-magnitude relation \citep[e.g.,][]{blakeslee2005} and
fundamental plane \citep[e.g.,][]{wuyts2004}, which suggests smooth
and regular evolution.  Moreover, neither in the field nor in clusters
has the early-type {\it fraction} changed significantly in
mass-selected samples \citep{holden2007,vanderwel2007}. Together with
these characteristics, the results presented in this paper fit into a
picture in which the field and cluster early-type galaxy populations
grow and change, but only while leaving many basic characteristics the
same.  In particular, we have demonstrated from the constancy of the
ellipticities of early-type galaxies over the last $\sim8$Gyr that the
bulge-to-disk ratio distribution of the cluster population
remains constant.  This suggests that processes that change the bulge-to-disk
ratio of individual early-type galaxies and the bulge-to-disk ratios of newly
formed or accreted early-type galaxies are balanced such that the
overall bulge-to-disk ratio distribution remains the same. It remains an open
question why this is the case, but it is clear that much can be
learned about the formation process of early-type galaxies by studying
their properties in even more detail and extending their observation
to higher redshifts.

\vspace{.3cm}

The authors would like to thank S. Adam Stanford for useful comments.
We would like to thank the anonymous referee for suggestions that
improved this paper.  Finally, we would like to thank Stefano Andreon
for pointing out previously published related papers, and Dave Wilman
for showing us his work ahead of publication.  ACS was developed under
NASA contract NAS5-32865, this research was supported by NASA grant
NAG5-7697.  We are grateful to K.~Anderson, J.~McCann, S.~Busching,
A.~Framarini, S.~Barkhouser, and T.~Allen for their invaluable
contributions to the ACS project at JHU.  This research has made use
of the NASA/IPAC Extragalactic Database (NED) which is operated by the
Jet Propulsion Laboratory, California Institute of Technology, under
contract with the National Aeronautics and Space Administration.  The
analysis pipeline used to reduce the DEIMOS data was developed at UC
Berkeley with support from NSF grant AST-0071048.


\appendix

\section{A.  Robustness of Ellipticity Measurements}
\label{ellrob}

We made a number of additional tests of the robustness of the
ellipticity measurements with the goal of estimating the magnitude of
any systematic errors.  These include adding fake galaxies into our
imaging data, comparing the ellipticity measurements made using
SExtractor to those using GALFIT, and using the a range of (incorrect)
PSFs to estimate the possible sensitivity to
systematic errors.

\subsection{A1. Simulations with Fake Galaxies}

We performed a number of simulations using fake galaxies to test
the reliability of the ellipticity measurements.  These simulations
were done in order to establish the best approach to use and to assess
how well we can measure the ellipticity, given our data.

The artificial galaxies were made by simulating two S{\'e}rsic models
with the same centroid, one with $n=4$ and one with $n=1$.  The two
components had the same luminosity, but different effective radii.  The
first effective radius was chosen to be three times the resolution of
the ACS camera, while the second was nine times the ACS resolution.  Each
component was given the same ellipticity.  These represent
well-resolved lenticular galaxies.  These fake images were then
simulated at magnitudes typical of the early-type galaxies in the
$z\sim 0.8-0.9$ clusters in our ACS imaging, and then added to a
real ACS image.  Reassuringly, the simulations recovered the input
ellipticities to a high accuracy, with a scatter of just $\sigma_e/e
=0.027$ and an offset of $\delta_e/e = -0.0007$.  The scatter and the
offset was highest for the most round objects.  For objects with an
input ellipticity of 0, the average recovered ellipticity was 0.03
with a scatter of $\sigma_e =0.014$.  As expected, objects that would
be seen as perfectly round given an arbitrarily high signal-to-noise,
are not measured to be round at typical signal-to-noise levels, but
the offset, as noted above, was quite small.

These simulations were made for a variety of bulge-to-disk ratios, by
varying the relative luminosities of the two components.  In addition,
we assumed that the intrinsic ellipticity of the $n=4$ bulge was
$\epsilon = 0.3$, or $\bar{\epsilon}=0.2$ when averaged over all
projections or over all $\cos(i)$.

\subsection{A2. Simulations with  Real Galaxies}
\label{simreal}

We compiled a number of images of low redshift elliptical and S0
galaxies.  We began with the catalog of \citet{frei96}.  That catalog
contains 24 elliptical and S0 galaxies, including two classified as
S0/a.  We expanded upon this by including 49 additional early-type
galaxies in the Virgo cluster.  Each was selected to be in the Virgo
cluster sample discussed in \citet{mei2007}.  Most of the galaxies in
that sample have accurate distances from surface brightness
fluctuations.  Those that do not have distances were assumed to be at
the center of the main Virgo concentration.

Once we selected the list of early-type galaxies, we extracted images
of those not in \citet{frei96} from the SDSS DR5 in the $g$ filter.
For those images from the SDSS, we removed nearby bright stars and
other galaxies, replacing the pixels with those randomly selected from
nearby blank regions.  This produced a final collection of 73 images
of nearby galaxies imaged in either $g$ or $B$.

We added these galaxies, appropriately scaled in size and
magnitude, to the imaging data used for each of the clusters, and
replicated the detection and measuring process. The galaxies were
resized as appropriate for the redshift of the cluster, convolved
with a representative PSF, and then noise was added
to each.  Each galaxy was simulated at a variety of input magnitudes,
and multiple realizations were done at each magnitude to evaluate the
effect of noise and binning.  This is the same process used in
\citet{holden2004}.  The magnitudes were selected to span the range of
observed galaxy magnitudes, to 0.4 mag below the limit used for
classifying galaxies in each sample. This is important to ensure that
the selection at the limit is not influenced by the magnitude cutoff.

\subsubsection{Recovery of the Input Galaxy's Ellipticity}

We compared the offset between the ellipticity measurement from the
original image and that of the simulated high redshift image.  In
general, the offset between the low redshift original image and high
redshift scaled image ellipticities are consistent.  For galaxies
within our magnitude limits, we found that the typical systematic
offset is from $\delta_{\epsilon} \sim -0.03$ to $\delta_{\epsilon}
\sim 0.01$, where the simulated galaxies are measured to be less
elliptical at high redshift than in their original, low redshift
images.  For galaxies at or below our selection limits
$\delta_{\epsilon} \sim -0.06$ to $\delta_{\epsilon} \sim 0.05$.  The
median systematic offset was $\delta_{\epsilon} \sim 0.01$ for our
whole sample of simulated galaxy images.  We show the ellipticities
measured in the output of the simulations as a function of the input
ellipticities in Figure \ref{cl_e}.  The scatter scales with the input
signal-to-noise as expected.  One exception is that, in general, the
measurements made with WFPC2 are worse than those made with ACS at the
same redshift.  This is expected, because WFPC2 is under-sampled and
the images are typically at a lower S/N.

\subsubsection{Random Errors}

We used the multiple realizations of the galaxy simulations to examine
the random errors.  The scatter in the measurements for a given input
galaxy in a given cluster observation is, on the whole, small.  The
ACS imaging has measurements with a median scatter of
$\sigma_{\epsilon}\sim 0.01$ in ellipticity, while the WFPC2 imaging
shows larger median scatter of $\sigma_{\epsilon} \sim 0.02 - 0.03$
in ellipticity.  These numbers are close to the typical errors
reported by GALFIT.  GALFIT's reported errors scale with the magnitude
of the galaxy, which we also find in our simulations.  At the faintest
magnitudes, at or below the magnitude limits for classification, we
find a scatter on the ellipticities as high as $\sigma_{\epsilon}
\sim 0.05$, once again matching the scatter in the ellipticity as
estimated by GALFIT.

\subsubsection{Systematic Errors from PSF Mismatches}
\label{psfbad}

Because the PSF so important, we performed
simulations where GALFIT was run using a different PSF than the PSF
that was used to convolve with the data.  We reran the simulations for
\ms\ using the correct \ia\ PSF when simulating the $z=0.83$ galaxy,
but we ran GALFIT using the \I\ PSF.  The differences between all
three PSFs in terms of encircled energy are small
\citep{sirianni2005}. However, there is an asymmetry in the PSF of the
\I\ filter \citep{sirianni2005,jee2007} that causes a noticeable
difference in the resulting PSFs.

\begin{figure}[htbp]
\begin{center}
\includegraphics[width=3.4in]{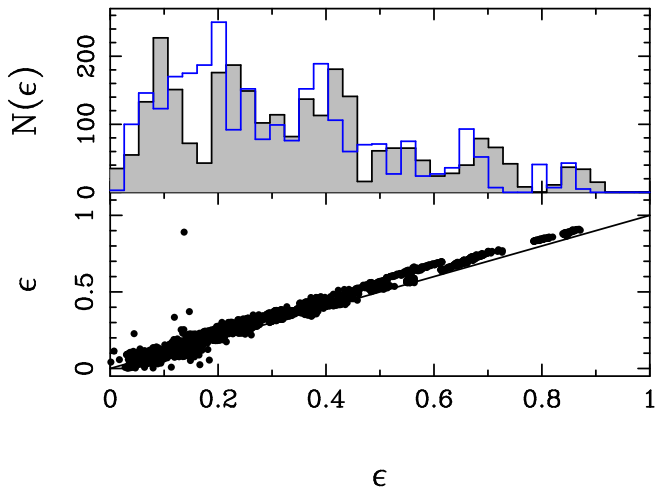}

\end{center}

\caption[f12.eps]{Ellipticity determined with wrong PSF versus
  ellipticity with the correct PSF. The bottom panel shows the output
  ellipticity as a function of the input ellipticity, with a line with
  a slope of one to guide the eye.  The top panel shows the input
  ellipticity distribution in blue and the output as the gray-shaded
  histogram.  The simulations were done for cluster galaxies at
  $z=0.83$ which were convolved with the \ia\ PSF, but the modeling of
  the galaxy luminosity profiles was done using the \I\ PSF has some
  asymmetry.  The use of the incorrect PSF in this case has only a
  modest impact, biasing the ellipticities towards higher values by
  $\delta_{\epsilon}\sim0.03$ at the highest ellipticities and
  becomes smaller, $\delta_{\epsilon}=0.018$ or a $\sim$6\% change,
  at the median input ellipticity.  }

\label{badpsf}
\end{figure}

We show the results of using the \I\ PSF to fit in Figure
\ref{badpsf}.  We find an overall offset of $\delta_{\epsilon}\sim
0.027$ between the average ellipticities, which is of order the same
size as the bins we use in the top panel.  Examining the lower panel,
we can see that the offset is not constant with ellipticity, but also
depends on the input galaxy used for the simulation.  For example,
along the x-axis at $\epsilon\sim 0.3$, there are a number of points
where the incorrect PSF yields $\epsilon\sim 0.4$ while other
galaxies have similar ellipticities when using the correct or
incorrect PSF.  From this we can conclude that typical systematic
errors on the ellipticity measurements are on the order of
$\delta_{\epsilon}\sim 0.01 - 0.03$, but this depends on the shape
of the galaxy.  The PSF is more asymmetric in the
redder passbands, so the systematic error is likely to be largest for
the highest redshift clusters.  The statistic we use the most is the
median ellipticity.  Using these simulations, we find that the offset
in the median ellipticity is $\delta_{\epsilon}=0.018$ for the \I\
data.  This is a quite a small effect, given the noticeable changes
that we have made to the PSF.

\subsection{A3. Comparison of Ellipticity Measures}
\label{simecomp}

Previous work, such as \citet{dressler97} and \citet{postman2005}, use
SExtractor's flux-weighted moments to estimate the ellipticities of
galaxies.  This approach, however, does not remove the effect of the
PSF.  Below we will compare the ellipticity
measurements we use with those used by previous work in the
literature.  We will also use our simulations to evaluate how well
these techniques reproduce the underlying ellipticity distribution.

\begin{figure*}[htbp]
\begin{center}
\includegraphics[width=6.5in]{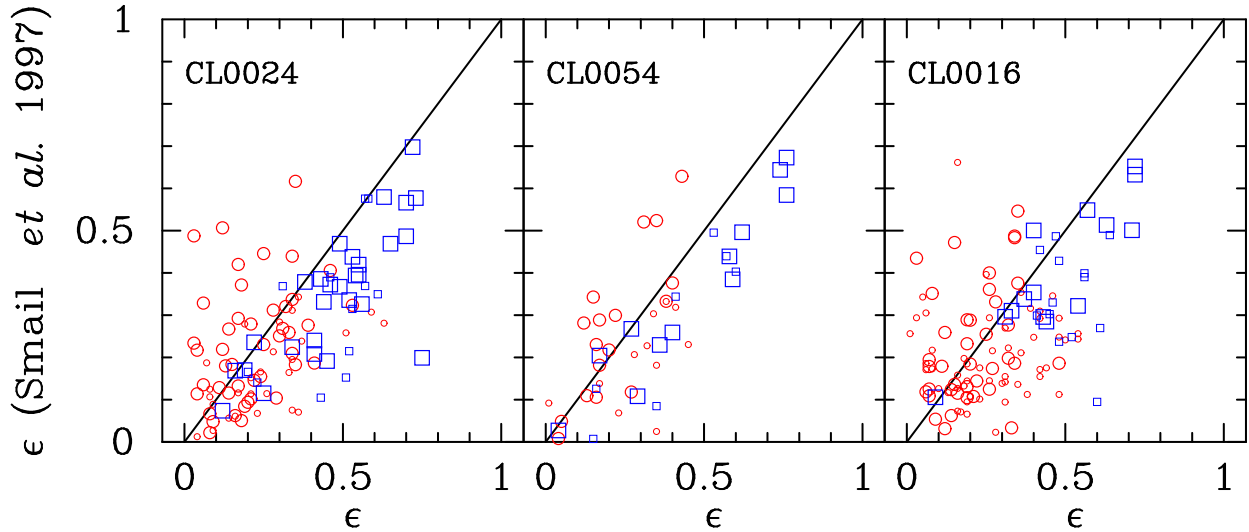}

\end{center}
\caption[f13.eps]{Ellipticity measurement comparison with
\citet{smail1997}.  Our ellipticities measured by fitting PSF convolved models to the images with GALFIT are on the x-axis.
The y-axis values are ellipticities from \citet{smail1997} measured
from the second-order flux-weighted moments of galaxies using
SExtractor.  Galaxies classified by eye as S0s are open blue squares
while ellipticals are red circles.  The smaller symbols are galaxies
with radii less than 0\farcs 3, or $\sim2$ WFPC2 resolution
elements.  The black line is a line of slope one with an intercept of
zero. Our ellipticities are generally less round than those of
\citet{smail1997}, and this is discrepancy is larger for the smaller
galaxies.  The median offset in the ellipticities for S0 galaxies is
$\delta \epsilon_{med} = -0.13$, much larger than the worst systematic
error we found in our simulations of our ellipticity measurements.
This highlights the importance of including the PSF in the measuring
process. }
\label{e_comp}
\end{figure*}

In Figure \ref{e_comp}, we compare our ellipticity measurements for
\koo\, \zwcl\ and \od\ with those from \citet{smail1997}.  Our
measurements, as we stated above, are based on fitting elliptical
models of S{\'e}rsic profiles to the data, including the effects of
the PSF.  The results from \citet{smail1997} are based on using
SExtractor \citep{bertin96}.  SExtractor estimates the ellipse by
computing the flux-weighted second-order moments of the galaxy.
SExtractor includes those pixels above the threshold defined by an
isophote as measured in a smoothed image.  \citet{smail1997} used a
detection isophote of $\sim 1.3 \sigma$ and a 0\farcs 3 diameter, or 3
pixels with WFPC2, top-hat filter for the smoothing.

We find that our ellipticities are less round than those from
\citet{smail1997}.  We hypothesize that this comes about for two
reasons.  First, our approach removes the smoothing from the
PSF which tends to circularize the ellipticity
measurements. Second, much of our data was observed with ACS, which
has a more compact PSF than WFPC2.  The median
offset in the ellipticities for S0 galaxies is $\delta \epsilon_{med}
= -0.13$.  This is a very large change, much larger than any of the
systematic error we found in our simulations of our ellipticity
measurements.  This highlights the importance of including the PSF in
the measuring process, and does indicate the challenge of using WFPC2
data for measurements at higher redshift.   

\begin{figure}[htbp]
\begin{center}
\includegraphics[width=4.5in]{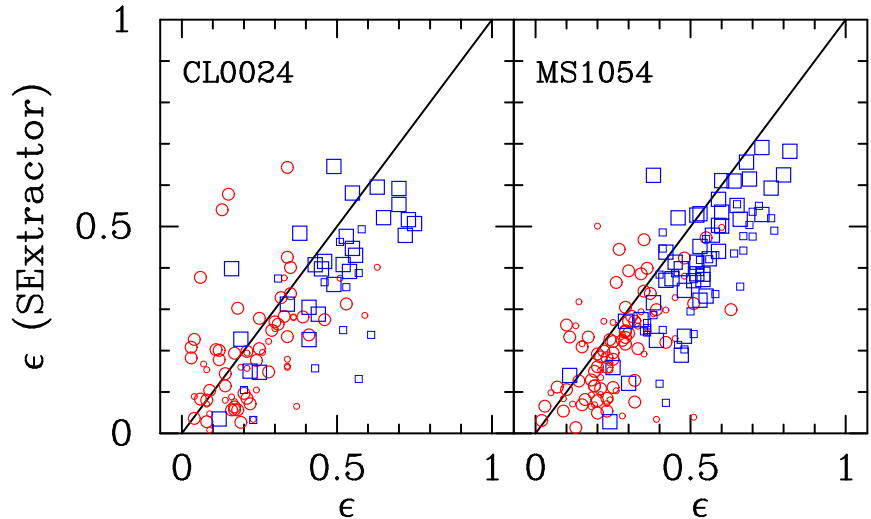}

\end{center}

\caption[f14.eps]{SExtractor measures of the ellipticity versus
those from GALFIT for ACS imaging data. This plot is similar to Fig.\
\ref{e_comp}, and uses the same symbols, but compares ellipticities
from measurements of the second-order flux-weighted moments made
with ACS imaging, as opposed to WFPC2 as in Fig.\ \ref{e_comp}, to
those from fitting models to the galaxies. As we see in Fig.\
\ref{e_comp}, the luminosity-weighted moment estimates are generally
rounder than the ones measured by GALFIT model fitting, despite the
higher resolution and better sampling of ACS as compared with WFPC2.
This figure shows that modeling the PSF when
measuring the ellipticity is particularly important for deriving a
robust estimate of the underlying value, and that the resolution of
the instrument is a lesser factor (at least in this case between
WFPC2 and ACS).  As in Fig.\ \ref{e_comp}, the galaxies classified as
S0s are more likely to have small ellipticities when they are larger
in size.  The median offset for S0 galaxies is $\delta \epsilon_{med}
= -0.10$, smaller than we found in Fig.\ \ref{e_comp}, but still much
larger than the statistical and systematic errors we find for our
PSF-corrected GALFIT model approach. }

\label{mye_comp}
\end{figure}

In Figure \ref{mye_comp}, we plot the ellipticities we measure
using SExtractor on the ACS images of \ms\ and \zwcl\ in comparison to
the ellipticities we find by fitting models using GALFIT.  From these
plots, we conclude that the more compact PSF for ACS
is not the dominate reason for the different ellipticities between our
work and the work of previous authors. Removing the effect of the PSF
is the most important component.

\begin{figure}[htbp]
\begin{center}
\includegraphics[width=6in]{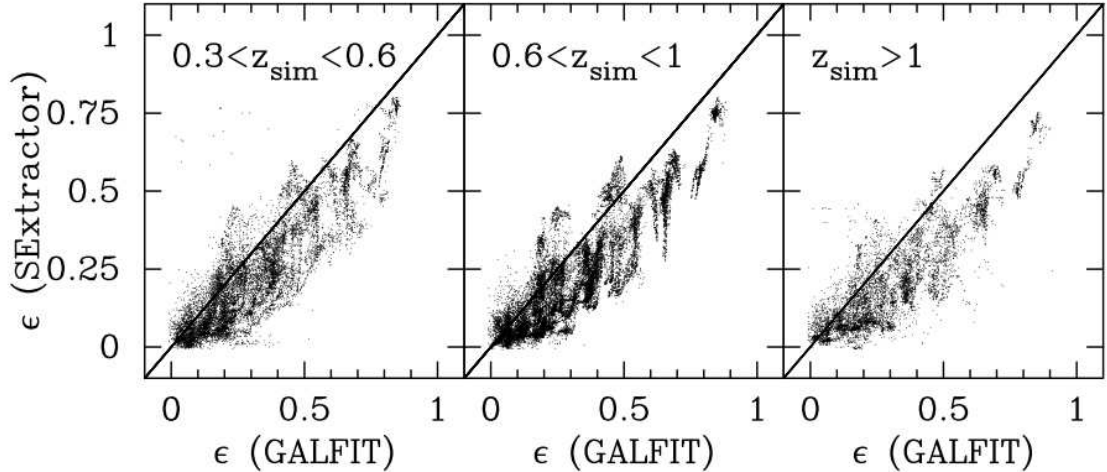}

\end{center}
\caption[f15.eps]{SExtractor ellipticities as a function of the GALFIT
  ellipticities for galaxies from our simulations.  We plot the
  SExtractor ellipticities on the y-axis as a function of the GALFIT
  ellipticities for the same simulated galaxies.  For comparison, we
  plot a line with a slope of one.  Note that this plot shows a
  similar behavior as Figs.\ \ref{e_comp} and \ref{mye_comp}, namely
  that the ellipticities as estimated by SExtractor are rounder than
  those from GALFIT.  The GALFIT measurements of the simulated $z>0.3$
  galaxies recover the actual ellipticities with an average offset of
  $\delta_{\epsilon} \sim -0.01$ (see Fig.\ \ref{cl_e}), much
  smaller than the offset of $\delta_{\epsilon} \sim -0.09$ we find
  between the GALFIT and SExtractor ellipticities.  The higher density
  of points in the $z=0.85$ redshift bin simply reflects the larger
  number of clusters included in that bin.}
\label{compare_e}
\end{figure}

Finally, we plot the comparison between the ellipticities as
derived by GALFIT and those from SExtractor in Figure \ref{compare_e}
for our simulations of high-redshift galaxies.  These are images of
real galaxies that have been redshifted and to which we have added
noise.  As with the previous plots showing trends in the data, the
simulated galaxies are measured to be rounder with SExtractor than by
fitting models using GALFIT.  The average offset is $\delta
\epsilon_{med} \sim -0.09$ between the SExtractor ellipticity and
the GALFIT ellipticity, in good agreement with the $\delta
\epsilon_{med} = -0.10$ we measure in Figure \ref{mye_comp}.  As we
show in Figure \ref{cl_e}, our GALFIT ellipticities reproduces the
underlying ellipticity distribution.  Because our GALFIT method
reproduces the input ellipticities so well, we conclude that the
SExtractor method of measuring ellipticities underestimates the
ellipticities of galaxies at higher redshifts. We would advise against
the use of SExtractor ellipticities for quantitative studies of high
redshift galaxies using ACS and WFPC2 where the PSF is a modest, but
significant fraction of the size of the object.  
 
\section{B. The Redshift Catalog for \ctw}
\label{ctw}

Of the clusters discussed here, only \ctw\ has no previously published
redshift catalog beyond the six galaxies listed in the discovery paper
\citep{ebeling2001}.  We have collected redshifts for 21
new members in the central region of the cluster, along with 10
more galaxies outside of the region covered by the ACS imaging.  30 of
these 31 galaxies were selected to have ACS \I\ magnitudes $<24$ mag AB
and colors in the range $1.5 < V_{606} - I_{814} < 2$, the remaining
one was a ``filler'' object with a bluer color.  The data were taken
using the DEIMOS spectrograph on Keck II, in a 1 hr exposure with
the 600 line mm$^{-1}$ grating in variable conditions.  We list the
catalog of redshifts in Table \ref{cl1226cat}.  

We measured redshifts by centroiding emission features and
cross-correlating with a variety of templates.  We computed the
biweight center and scale of all galaxies with redshifts within $0.87
< z < 0.91$.  We found $\bar{z} = 0.890 \pm 0.001$ with a dispersion
of 1143 $\pm$ 162 km s$^{-1}$, in good agreement with the 1270 km
s$^{-1}$ quoted in \citet{jorgensen2006}.  We used a biweight to
estimate both the redshift and the velocity dispersion of the cluster
with errors from the jackknife of the galaxy redshifts.  The redshift
range we used for membership excludes two galaxies at $z\sim0.91$.
Including those two objects raises the dispersion to 1322 $\pm$ 221 km
s$^{-1}$.  It is often the case that, with small numbers of galaxies
that the dispersion is overestimated -- see for example the
dispersions quoted for \hogb\ and \hoga\ in \citet{lubin2000} versus
those from the much larger sample in \citet{gal2005}.  This arises
because nearby large scale structure or groups of galaxies can project
into line of sight of the cluster.  We opted to remove the two
galaxies at $z\sim0.91$ to provide conservative estimate of the
dispersion, which we also list in Table \ref{summary}.

\begin{deluxetable*}{lrrrrr}
\tablecolumns{6}
\tablecaption{Catalog for \ctw}
\tablehead{\colhead{Obj} & \colhead{R.A.} & 
\colhead{Dec.} & \colhead{I}
 & \colhead{V-I} & \colhead{z}
  \\
\colhead{} & \colhead{(J2000)} & \colhead{(J2000)} & \colhead{(mag
  AB)} & \colhead{mag AB} & \colhead{}  \\}

\startdata
GAL12263974+3329315 & 12:26:39.74 & 33:29:31.51 & 21.82 $\pm$ 0.02  & 1.75 $\pm$ 0.03  & 1.035 \\
GAL12264240+3330094 & 12:26:42.40 & 33:30:09.40 & 22.11 $\pm$ 0.01  & 1.67 $\pm$ 0.01  & 0.929 \\
GAL12263338+3330277 & 12:26:33.38 & 33:30:27.77 & 21.68 $\pm$ 0.01  & 1.82 $\pm$ 0.02  & 0.911 \\
GAL12264935+3330312 & 12:26:49.35 & 33:30:31.25 & 21.61 $\pm$ 0.01  & 1.82 $\pm$ 0.01  & 0.891 \\
GAL12263934+3330353 & 12:26:39.34 & 33:30:35.35 & 22.40 $\pm$ 0.01  & 1.96 $\pm$ 0.03  & 0.943 \\
GAL12264600+3330347 & 12:26:46.00 & 33:30:34.74 & 21.43 $\pm$ 0.01  & 1.84 $\pm$ 0.01  & 0.892 \\
GAL12263157+3330473 & 12:26:31.57 & 33:30:47.38 & 22.99 $\pm$ 0.02  & 0.72 $\pm$ 0.02  & 0.963 \\
GAL12262950+3330580 & 12:26:29.50 & 33:30:58.01 & 21.59 $\pm$ 0.01  & 1.81 $\pm$ 0.02  & 1.211 \\
GAL12264172+3331041 & 12:26:41.72 & 33:31:04.15 & 22.26 $\pm$ 0.01  & 0.59 $\pm$ 0.01  & 0.690 \\
GAL12264671+3331037 & 12:26:46.71 & 33:31:03.73 & 22.07 $\pm$ 0.01  & 1.60 $\pm$ 0.01  & 0.767 \\
GAL12263998+3331047 & 12:26:39.98 & 33:31:04.72 & 21.70 $\pm$ 0.01  & 1.85 $\pm$ 0.02  & 0.929 \\
GAL12263648+3331124 & 12:26:36.48 & 33:31:12.47 & 22.47 $\pm$ 0.02  & 0.99 $\pm$ 0.02  & 0.846 \\
GAL12263107+3331349 & 12:26:31.07 & 33:31:34.97 & 22.73 $\pm$ 0.01  & 0.60 $\pm$ 0.01  & 0.965 \\
GAL12263402+3331356 & 12:26:34.02 & 33:31:35.62 & 21.84 $\pm$ 0.01  & 1.84 $\pm$ 0.02  & 0.966 \\
GAL12265120+3331385 & 12:26:51.20 & 33:31:38.54 & 22.68 $\pm$ 0.01  & 1.90 $\pm$ 0.02  & 0.881 \\
GAL12265293+3331465 & 12:26:52.93 & 33:31:46.54 & 22.78 $\pm$ 0.01  & 1.76 $\pm$ 0.02  & 0.964 \\
GAL12263832+3331482 & 12:26:38.32 & 33:31:48.21 & 23.27 $\pm$ 0.06  & 0.63 $\pm$ 0.09  & 0.542 \\
GAL12262701+3331580 & 12:26:27.01 & 33:31:58.09 & 22.27 $\pm$ 0.08  & 1.82 $\pm$ 0.08  & 0.645 \\
GAL12272134+3332083 & 12:27:21.34 & 33:32:08.32 & 21.86 $\pm$ 0.01  & 1.81 $\pm$ 0.01  & 0.893 \\
GAL12262817+3332232 & 12:26:28.17 & 33:32:23.23 & 22.93 $\pm$ 0.02  & 1.13 $\pm$ 0.03  & 0.673 \\
GAL12271941+3332269 & 12:27:19.41 & 33:32:26.91 & 23.08 $\pm$ 0.02  & 0.70 $\pm$ 0.02  & 0.885 \\
GAL12265525+3332324 & 12:26:55.25 & 33:32:32.43 & 22.28 $\pm$ 0.01  & 1.90 $\pm$ 0.01  & 0.895 \\
GAL12271460+3332377 & 12:27:14.60 & 33:32:37.70 & 22.10 $\pm$ 0.01  & 1.89 $\pm$ 0.01  & 0.881 \\
GAL12265629+3332414 & 12:26:56.29 & 33:32:41.48 & 22.53 $\pm$ 0.01  & 1.78 $\pm$ 0.01  & 0.893 \\
GAL12265995+3332405 & 12:26:59.95 & 33:32:40.54 & 22.48 $\pm$ 0.01  & 1.76 $\pm$ 0.02  & 0.892 \\
GAL12265923+3332405 & 12:26:59.23 & 33:32:40.59 & 22.21 $\pm$ 0.01  & 1.82 $\pm$ 0.01  & 0.897 \\
GAL12265689+3332437 & 12:26:56.89 & 33:32:43.75 & 22.86 $\pm$ 0.01  & 1.84 $\pm$ 0.02  & 0.896 \\
GAL12270510+3332475 & 12:27:05.10 & 33:32:47.53 & 23.01 $\pm$ 0.01  & 0.77 $\pm$ 0.01  & 0.798 \\
GAL12265060+3332461 & 12:26:50.60 & 33:32:46.18 & 21.50 $\pm$ 0.01  & 1.83 $\pm$ 0.01  & 0.875 \\
GAL12265825+3332485 & 12:26:58.25 & 33:32:48.57 & 19.13 $\pm$ 0.00  & 1.96 $\pm$ 0.00  & 0.891 \\
GAL12271547+3332539 & 12:27:15.47 & 33:32:53.91 & 22.87 $\pm$ 0.02  & 0.69 $\pm$ 0.02  & 1.034 \\
GAL12270083+3333019 & 12:27:00.83 & 33:33:01.97 & 22.50 $\pm$ 0.01  & 1.35 $\pm$ 0.02  & 0.914 \\
GAL12265214+3333071 & 12:26:52.14 & 33:33:07.12 & 23.02 $\pm$ 0.02  & 1.76 $\pm$ 0.02  & 0.881 \\
GAL12265714+3333046 & 12:26:57.14 & 33:33:04.61 & 21.65 $\pm$ 0.01  & 1.95 $\pm$ 0.02  & 0.930 \\
GAL12270134+3333044 & 12:27:01.34 & 33:33:04.42 & 21.27 $\pm$ 0.00  & 1.89 $\pm$ 0.01  & 0.897 \\
GAL12264489+3333094 & 12:26:44.89 & 33:33:09.46 & 22.95 $\pm$ 0.01  & 1.83 $\pm$ 0.02  & 1.196 \\
GAL12271887+3333137 & 12:27:18.87 & 33:33:13.72 & 21.73 $\pm$ 0.01  & 1.85 $\pm$ 0.01  & 0.897 \\
GAL12271572+3333111 & 12:27:15.72 & 33:33:11.11 & 21.60 $\pm$ 0.01  & 1.73 $\pm$ 0.01  & 0.892 \\
GAL12270766+3333138 & 12:27:07.66 & 33:33:13.82 & 22.61 $\pm$ 0.01  & 1.65 $\pm$ 0.02  & 0.902 \\
GAL12272080+3333163 & 12:27:20.80 & 33:33:16.37 & 21.60 $\pm$ 0.01  & 1.83 $\pm$ 0.01  & 0.819 \\
GAL12271607+3333192 & 12:27:16.07 & 33:33:19.25 & 22.47 $\pm$ 0.02  & 1.85 $\pm$ 0.02  & 0.901 \\
GAL12264296+3333195 & 12:26:42.96 & 33:33:19.59 & 22.77 $\pm$ 0.01  & 1.00 $\pm$ 0.02  & 0.849 \\
GAL12265244+3333239 & 12:26:52.44 & 33:33:23.90 & 22.22 $\pm$ 0.02  & 1.86 $\pm$ 0.02  & 0.893 \\
GAL12270671+3333266 & 12:27:06.71 & 33:33:26.65 & 21.87 $\pm$ 0.01  & 1.72 $\pm$ 0.01  & 0.883 \\
GAL12271695+3333261 & 12:27:16.95 & 33:33:26.19 & 21.45 $\pm$ 0.01  & 1.87 $\pm$ 0.02  & 0.891 \\
GAL12265312+3333310 & 12:26:53.12 & 33:33:31.08 & 21.90 $\pm$ 0.01  & 1.87 $\pm$ 0.02  & 0.896 \\
GAL12264799+3333348 & 12:26:47.99 & 33:33:34.80 & 22.29 $\pm$ 0.01  & 0.78 $\pm$ 0.01  & 0.330 \\
GAL12270750+3333439 & 12:27:07.50 & 33:33:43.96 & 22.62 $\pm$ 0.01  & 0.90 $\pm$ 0.01  & 0.850 \\
GAL12271102+3333471 & 12:27:11.02 & 33:33:47.10 & 21.71 $\pm$ 0.01  & 1.80 $\pm$ 0.01  & 0.755 \\
GAL12270440+3334063 & 12:27:04.40 & 33:34:06.35 & 22.74 $\pm$ 0.01  & 1.82 $\pm$ 0.02  & 0.887 \\
GAL12270214+3334060 & 12:27:02.14 & 33:34:06.06 & 21.48 $\pm$ 0.01  & 1.61 $\pm$ 0.02  & 0.896 \\
GAL12271313+3334130 & 12:27:13.13 & 33:34:13.02 & 21.63 $\pm$ 0.01  & 1.78 $\pm$ 0.01  & 0.763 \\
GAL12271082+3334195 & 12:27:10.82 & 33:34:19.58 & 22.40 $\pm$ 0.01  & 1.06 $\pm$ 0.02  & 0.767 \\
GAL12270730+3334237 & 12:27:07.30 & 33:34:23.78 & 22.07 $\pm$ 0.01  & 1.82 $\pm$ 0.02  & 0.892 \\
GAL12271975+3334272 & 12:27:19.75 & 33:34:27.26 & 21.30 $\pm$ 0.01  & 0.80 $\pm$ 0.01  & 0.530 \\
GAL12272074+3334437 & 12:27:20.74 & 33:34:43.71 & 21.30 $\pm$ 0.01  & 1.79 $\pm$ 0.01  & 0.762 \\
GAL12271545+3334558 & 12:27:15.45 & 33:34:55.82 & 21.68 $\pm$ 0.01  & 1.61 $\pm$ 0.01  & 0.768 \\
GAL12265239+3335006 & 12:26:52.39 & 33:35:00.60 & 22.75 $\pm$ 0.01  & 1.67 $\pm$ 0.02  & 0.764 \\
GAL12270219+3335182 & 12:27:02.19 & 33:35:18.22 & 22.64 $\pm$ 0.02  & 1.55 $\pm$ 0.02  & 0.891 \\
GAL12270676+3335289 & 12:27:06.76 & 33:35:28.92 & 21.58 $\pm$ 0.01  & 1.77 $\pm$ 0.01  & 0.755 \\
\enddata
\label{cl1226cat}
\end{deluxetable*}

\newpage


\end{document}